\documentclass[11pt]{article}
\newcommand{\Comment}[1]{{}}
\usepackage{amsfonts,amsthm,amsmath,amssymb,slashed}
\usepackage[textwidth = 430 pt, textheight = 630 pt]{geometry}
\usepackage{color}

\Comment{\usepackage{color}
\definecolor{MyDarkBlue}{rgb}{0.15,0.15,0.45}
\usepackage[linktocpage=true]{hyperref}
\hypersetup{
colorlinks=true,
citecolor=MyDarkBlue,
linkcolor=MyDarkBlue,
urlcolor=MyDarkBlue,
pdfauthor={Horatiu Nastase and Jacob Sonnenschein},
pdftitle={The title},
pdfsubject={hep-th}
}

\usepackage[numbers,sort&compress]{natbib}
\usepackage{hypernat}}
\usepackage{graphicx}
\usepackage{cite}

\newcommand\ignore[1]{}
\def\one{{\,\hbox{1\kern-.8mm l}}}

\def\Tr{{\rm Tr\, }}

\def\a{\alpha}\def\b{\beta}

\def\d{\partial}

\def\Tr{\mathop{\rm Tr}\nolimits}

\newcommand{\Cset}{{\,\,{{{^{_{\pmb{\mid}}}}\kern-.45em{\mathrm C}}}}}

\newcommand{\be}{\begin{equation}}
\newcommand{\bea}{\begin{eqnarray}}

\newcommand{\ee}{\end{equation}}
\newcommand{\eea}{\end{eqnarray}}

\begin{document}

\renewcommand{\thefootnote}{\fnsymbol{footnote}}

\makeatletter
\@addtoreset{equation}{section}
\makeatother
\renewcommand{\theequation}{\thesection.\arabic{equation}}

\rightline{}
\rightline{}




\begin{center}
{\LARGE \bf{\sc $T\bar T$ deformations and the pp wave correspondence}}
\end{center} 
 \vspace{1truecm}
\thispagestyle{empty} \centerline{
{\large \bf {\sc Horatiu Nastase${}^{a}$}}\footnote{E-mail address: \Comment{\href{mailto:horatiu.nastase@unesp.br}}
{\tt horatiu.nastase@unesp.br}}
{\bf{\sc and}}
{\large \bf {\sc Jacob Sonnenschein${}^{b,c}$}}\footnote{E-mail address: \Comment{\href{mailto:cobi@tauex.tau.ac.il}}{\tt cobi@tauex.tau.ac.il}}
                                                        }

\vspace{.5cm}


\centerline{{\it ${}^a$Instituto de F\'{i}sica Te\'{o}rica, UNESP-Universidade Estadual Paulista}} 
\centerline{{\it R. Dr. Bento T. Ferraz 271, Bl. II, Sao Paulo 01140-070, SP, Brazil}}
\vspace{.3cm}
\centerline{{\it ${}^b$School of Physics and Astronomy,}}
\centerline{{\it The Raymond and Beverly Sackler Faculty of Exact Sciences, }} \centerline{{\it Tel Aviv University, Ramat Aviv 69978, Israel}}
\vspace{.3cm}
\centerline{{\it ${}^c$ Simons Center for Geometry and Physics,}} 
\centerline{{\it SUNY, Stony Brook, NY 11794, USA }}

\vspace{1truecm}

\thispagestyle{empty}

\centerline{\sc Abstract}

\vspace{.4truecm}

\begin{center}
\begin{minipage}[c]{380pt}
{\noindent In this paper we consider $T\bar T$ deformations in the context of pp waves obtained 
from gravity duals. We propose a deformation of $AdS_5\times S^5$ similar to the deformation of the 
single trace $T\bar T$ deformation of $AdS_3\times S^3\times T^4$ with NS-NS flux, and study it 
through the Penrose limit, concluding that it must correspond to some dipole theory, probably
noncommutative. We $T\bar T$ deform the worldsheet string for the $AdS_5\times S^5$ pp wave, 
and find a corresponding spin chain Hamiltonian. Finally, directly $T\bar T$ deforming the spin chain
Hamiltonian obtained from the pp wave, we find that it  corresponds to  an 
equivalent BMN sector of the ${\cal N}=4$ SYM.

}
\end{minipage}
\end{center}

\vspace{.5cm}

\setcounter{page}{0}
\setcounter{tocdepth}{2}

\newpage

\tableofcontents
\renewcommand{\thefootnote}{\arabic{footnote}}
\setcounter{footnote}{0}

\linespread{1.1}
\parskip 4pt



\section{Introduction}

The $T\bar T$ deformations   of quantum field theories in two dimensions, defined by 
Zamolodchikov in \cite{Zamolodchikov:2004ce,Smirnov:2016lqw}, which are of special importance for
 integrable theories, have attracted a lot of interest recently. Although originally defined 
as deformations of  renormalized theories, through the normal ordered product of the 
energy-momentum tensors $T$ and $\bar T$ (more precisely, the $\det T_{\mu\nu}$ operator, 
so $\frac{1}{8}(T_{\a\b}T^{\a\b}-({T^\a}_\a)^2)$), 
in point splitting regularization, it was soon realized that
one can equivalently define the theory by deforming the classical Lagrangian density by $\det T_{\mu\nu}$ at 
each step of the deformation, thus finding closed forms for the Lagrangian, at least in the scalar case
\cite{Cavaglia:2016oda,Bonelli:2018kik}.\footnote{Note that there are arguments \cite{Barbon:2020amo} that the 
$T\bar T$ deformations are generically thermodynamically unstable at high temperature.} 
This also leads to deformations of the classical solutions of the theory, see
\cite{Conti:2018jho,Conti:2018tca,Nastase:2020evb,Nastase:2021mgy}.
In higher dimensions, the equivalent of the $T\bar T$ deformations is less understood, 
and there are  several proposals of deformations\cite{Bonelli:2018kik,Taylor:2018xcy,Babaei-Aghbolagh:2020kjg} 
(although one can also simply extend 
the $T\bar T$ deformed actions to higher dimensions \cite{Nastase:2021mgy}). 

Given that the most interesting applications are to conformal and integrable field theories, a natural
question is, can one define a gravity dual that corresponds to the $T\bar T$ deformations? 
A first such proposal was considered in \cite{McGough:2016lol}, though it just stated that the 
deformation amount to giving Dirichlet boundary conditions at a finite cut-off position from 
the original boundary, in the RG direction, $r=r_c$. This, however, is not quite the definition one 
would like, namely to have a deformed gravity dual of the deformed boundary field theory. 

That has been obtained, in a certain sense, in the case of string theory in the $AdS_3\times S^3\times 
T^4$ with NS-NS flux, considered in \cite{Giveon:2017nie,Giveon:2017myj}, with a boundary theory given
 by \({\cal M}^p/ \mathfrak{S}_p\), where \({\cal M}\) is a CFT with central charge \(c=6k\), and \(k\) is the number 
 of NS5 branes,  \(p\) is the number of fundamental strings, with the duality to the symmetric product space known to be
 valid for $k=1$. In that case,
one can construct string worldsheet vertex operators that correspond to the operators in the boundary 
CFT that are ``single trace'', $\sum_{i=1}^p {\cal O}_i$, where ${\cal O}_i$ is an ${\cal O}(x)\in 
{\cal M}$ that lives in the $i$th factor in the CFT product space.  Then  the 
usual $T\bar T$ deformation of the CFT would be of a ``double trace type'', namely the product 
of $T(x)$ and $\bar T(x)$, both of which are single traces $\sum_{i=1}^p T_i(x)$ in the CFT, 
and each have a corresponding 
string worldsheet (integrated) vertex operator for the bulk theory. Yet what is easier to 
understand in the gravity dual is instead the ``single trace'' deformation, 
\be
D(z,\bar z)=A\sum_{i=1}^p T_i(z)\bar T_i(\bar z)\;,
\ee
that has a corresponding (integrated) vertex operator for the bulk string worldsheet, and 
it has been shown \cite{Araujo:2018rho} to be given by a TsT transformation of the bulk in the $CFT_2$
directions $x$ and $t$, and to have many of the properties of the usual $T\bar T$ deformation of
$CFT_2$. 

We want then first to ask: can we generalize this construction to the $CFT_4$ case, i.e., to the 
$AdS_5\times S^5$ background? We would take the agnostic point of view, and construct a gravity 
dual, then try to understand the deformation of ${\cal N}=4$ SYM it corresponds to, by taking 
a Penrose limit, which  usually simplifies a lot the analysis \cite{Berenstein:2002jq}.

Reversely, the second question we ask is: what is the $T\bar T$ deformation of the string 
worldsheet for the $AdS_5\times S^5$ (maximally supersymmetric) pp wave, and what does it 
correspond for the spin chain in ${\cal N}=4$ SYM? We can either deform the string worldsheet theory, 
then discretize it to obtain a corresponding spin chain Hamiltonian, or consider the deformation of the 
original pp wave spin chain Hamiltonian directly, employing a $T\bar T$ deformation procedure for 
one dimensional (quantum mechanics) Hamiltonians, which we do using the procedure defined in 
\cite{Gross:2019ach}. 

The paper is organized as follows. In section 2 we review the 2 dimensional case, for the gravity dual to
the single trace $T\bar T$ deformation, and its TsT construction. In section 3 we apply the procedure
to the 4 dimensional case. In section 4 we take the Penrose limit of the gravity duals to the 
single trace deformation for $AdS_3\times S^3\times T^4$, and then to the proposed $AdS_5\times S^5$
deformation, and finally propose an interpretation for the deformation in ${\cal N}=4$ SYM. 
In section 5, we consider the $T\bar T$ deformation of the maximally supersymmetric pp wave 
obtained in the Penrose limit of $AdS_5\times S^5$, first as a deformation of the string worldsheet, 
discretized to give us a deformed spin chain Hamiltonian, and then finally as a deformation of the 
quantum mechanical spin chain Hamiltonian (obtained from the pp wave) itself, and propose a 
dual interpretation for this spin chain from ${\cal N}=4$ SYM. In section 6  we summarize and list 
open questions. In appendix A we review the derivation of the $T\bar T$ deformation of a
general quantum mechanical model.

\section{TsT transformations and $T\bar T$ deformations: two-dimensional case}

In the context of the AdS/CFT correspondence, in particular in the $AdS_3/CFT_2$ case, there is 
a proposed relation 
between the $T\bar T$ deformation of the $CFT_2$ at the boundary and a "single-trace" $T\bar T$ deformation on 
the string worldsheet in the bulk \cite{Giveon:2017nie,Giveon:2017myj}, of the type
\be 
T(z)\bar{T}(\bar{z})\equiv A \sum_i T_i(z) \bar{T}_i(\bar{z})\; .
\ee

Specifically, considering the $AdS_3\times S^3\times T^4$ solution with NS-NS flux,
(using the notation in \cite{Araujo:2018rho})
\bea
R^{-2}ds^2&=& e^{2\rho}(-dt^2+dx^2)+d\rho^2+\frac{1}{4}(\sigma_1^2+\sigma_2^2+\sigma_3^2)+ds^2(T^4)\cr
H&=& -2 e^{2\rho}dt\wedge dx\wedge d\rho+\frac{1}{4}\sigma_1\wedge \sigma_2\wedge \sigma_3\;,
\eea
and a constant dilaton $\Phi=\Phi_0$, the boundary CFT is given
by \({\cal M}^p/ \mathfrak{S}_p\), where \({\cal M}\) is a CFT with central charge \(c=6k\), and \(k\) 
is the number of NS5 branes and \(p\) is the number of fundamental strings (we restrict to $k=1$ if we want to 
be sure that we have 
the symmetric product boundary CFT). In the $B_{\rho 0}=B_{\rho 1}=0$ gauge, we have $B_{01}=-e^{2\rho}$.

Note that in this paper we will restrict to solutions with NS-NS flux; the case of R-R flux is not known to be related 
to a single-trace $T\bar T$ deformation of a tensor product CFT, hence we will not analyze it; its pp wave limit will 
also be considerably different due to the different fluxes. 

Then, as observed in \cite{Araujo:2018rho}, the $T\bar T$ deformed solution (corresponding to the single 
trace $T\bar T$ on the worldsheet)
in the notation in \cite{Chakraborty:2020udr} (the volume of $T^4$ is $(2\pi l_s)^4v$),
\bea
\frac{ds^2}{l_s^2}&=& \frac{k(-dt^2+dx^2)}{\frac{l_s^2}{R^2}+e^{-2\phi}}+kd\phi^2+ kds^2_{S^3}+ds^2_{T^4}\cr
e^{2\Phi}&=& \frac{vk}{p}\frac{e^{-2\phi}}{\frac{l_s^2}{R^2}+e^{-2\phi}}\equiv 
e^{2\Phi_0}\frac{e^{-2\phi}}{\frac{l_s^2}{R^2}+e^{-2\phi}}\;,\cr
H&=& -\frac{2e^{2\phi}}{\left(1+\frac{l_s^2}{R^2} e^{2\phi}\right)^2}dt\wedge dx\wedge d\phi
+\frac{k l_s^2}{4}\sigma_1\wedge \sigma_2\wedge \sigma_3\label{dilaton}
\eea
can be obtained as a TsT transformation in the $CFT_2$ directions $x$ and $t$ (though there is 
no proof of the relation to TsT outside this specific case).  

To see that, write the TsT transformations from \cite{Frolov:2005dj}, with $-2\gamma=l_s^2/R^2$
(note that the overall scale of the metric is $R_{\rm AdS}=kl_s^2$). 
We have a T-duality on $\phi_1$ (isometry 
direction), then $\phi_2\rightarrow \phi_2+\gamma\phi_1$, then T-duality back on $\phi_1$. In the shift step, this changes the 
metric as 
\bea
ds^2&=&\hat g_{11}d\phi_1^2+\hat g_{22}d\phi_2^2+2\hat g_{12}d\phi_1d\phi_2+2\hat g_{23} d\phi_2d\phi_3+2\hat g_{24}d\phi_2d\phi_4
+...\rightarrow\cr
ds^2&=& \hat g_{11}d\phi_1^2+\hat g_{22}(d\phi_2+\gamma d\phi_1)^2+2\hat g_{12}d\phi_1(d\phi_2+\gamma d\phi_1)+
2\hat g_{23}(d\phi_2+\gamma d\phi_1)d\phi_3+...\Rightarrow\cr
\hat G_{11}&=&\hat g_{11}+\gamma^2 \hat g_{22}+2\gamma \hat g_{12}\cr
\hat G_{1i}&=&\hat g_{1i}+\gamma \hat g_{2i}\;,\; \forall i\neq 1\;,
\eea
and similar relations for other fields (if there are any). On the other hand, for T-duality, the Buscher rules are 
\bea
\tilde G_{11}&=& \frac{1}{G_{11}}\;,\;\; \tilde G_{1i}=\frac{B_{1i}}{G_{11}}\cr
\tilde G_{ij}&=& G_{ij}-\frac{G_{1i}G_{1j}-B_{1i}B_{1j}}{G_{11}}\cr
\tilde B_{1i}&=& \frac{G_{1i}}{G_{11}}\cr
\tilde B_{ij}&=& B_{ij}-\frac{G_{1i}B_{1j}-B_{1i}G_{1j}}{G_{11}}\cr
\tilde \Phi&=& \Phi-\frac{1}{2}\log G_{11}.
\eea

Then, indeed: after a T-duality in time $t$, we get (dropping the $R$ factors)
\bea
ds^2&=& -e^{-2\rho}dt^2+2dtdx+d\rho^2+\frac{1}{4}(\sigma_1^2+\sigma_2^2+\sigma_3^2)+ds^2(T^4)\cr
\Phi&=&\Phi_0-\frac{1}{2}\log e^{2\rho}\;,\cr
B_{01}&=&0.
\eea

After the shift $x\rightarrow x+\gamma t$, we find 
\bea
ds^2&=& dt^2(-e^{-2\rho}+2\gamma)+2dtdx+d\rho^2+\frac{1}{4}(\sigma_1^2+\sigma_2^2+\sigma_3^2)+ds^2(T^4)\cr
\Phi&=&\Phi_0-\frac{1}{2}\log e^{2\rho}\;,\cr
B_{01}&=&0.
\eea

And after the T-duality back on $t$, we find 
\bea
ds^2&=& \frac{e^{2\rho}(-dt^2+dx^2)}{1-2\gamma e^{2\rho}}+d\rho^2+\frac{1}{4}(\sigma_1^2+\sigma_2^2+\sigma_3^2)+ds^2(T^4)\cr
\Phi&=& \Phi_0-\frac{1}{2}\log e^{2\rho}-\frac{1}{2}\log (e^{-2\rho}-2\gamma)\Rightarrow e^{2\Phi}=e^{2\Phi_0}\frac{e^{-2\rho}}{e^{-2\rho}
-2\gamma}\cr
B_{01}&=& -\frac{e^{2\rho}}{1-2\gamma e^{2\rho}}\Rightarrow H_{01\rho}=-\frac{2e^{2\rho}}{(1-2\gamma e^{2\rho})^2}.
\eea

The TsT background above is an example of a Yang-Baxter transformation. It corresponds to the classical r-matrix
\be 
r=\frac{1}{2} P_0\wedge P_1\;,
\ee
where \(P_0\) and \(P_1\) are generators of the Poincar\'{e} group. In general
\be 
r= A \wedge B\;,
\ee
where we say that the transformation is said to be \emph{abelian} if \([A,B]=0\). 
See~\cite{Araujo:2017jap} to see how we can build the deformed backgrounds. 

For a background of the form \(AdS_d\times M_{10-d}\), TsT deformations of the inner 
space \(M_{10-d}\) give marginal deformations of the \(CFT_{d-1}\), while TsT deformations of the AdS space give massive theories. 

In the \(AdS_3\times \mathbb{S}^3\times T^4\) case, the Yang-Baxter deformation \(r=\frac{1}{2} P_0\wedge P_1\) 
gives the dual of the \(T\bar{T}\)-deformation.

\section{TsT transformations and $T\bar T$ deformations: four-dimensional case}

In dimensions higher than two, the definition of $T\bar T$ deformations is not unique: there are several proposals.

The first paper to 
deal with this, \cite{Bonelli:2018kik}, notes that the 2-dimensional deformation
\be
\d_t S=\frac{1}{2}\int d^2x \sqrt{g}\left[\left(\epsilon^{\mu\nu}\epsilon^{\rho\sigma}\right)T_{\mu\rho}T_{\nu\sigma}\right]
=\int d^2x \sqrt{g} \det T_{\mu\nu}
\ee
can be generalized by noting that in 2 dimensions 
$\epsilon^{\mu\nu}\epsilon^{\rho\sigma}=g^{\mu\rho}g^{\nu\sigma}-g^{\nu\rho}g^{\mu\sigma}$, 
and then the above deformation can be trivially generalized to any dimension to the quadratic form
\be
\d_t S=\frac{1}{2}\int d^2x \sqrt{g}\left[\left(g^{\mu\rho}g^{\nu\sigma}-g^{\nu\rho}g^{\mu\sigma}\right)T_{\mu\rho}T_{\nu\sigma}\right]\;,
\ee
which now however is not a determinant anymore. They, however, do not explore this further, other than to say that for a single 
scalar field (with $X=(\d_\mu\phi)^2$), one gets
\be
\d_t{\cal L}=(D-1)\left[\frac{D}{2}{\cal L}^2-2X\d_X{\cal L} {\cal L}\right].
\ee
 
Instead, they consider the generalization with a determinant to an arbitrary power $\a$, $(-\det T)^{\frac{1}{\a}}$, 
\be
\d_t S=\frac{1}{\a-D}\int \sqrt{g}\left[-\frac{1}{D!}\epsilon^{\mu_1...\mu_D}\epsilon^{\nu_1...\nu_D} T_{\mu_1\nu_1}...T_{\mu_D\nu_D}\right]
^{\frac{1}{\a}}\;,
\ee
and find that they can write a closed form expression of the deformed Lagrangian of a free scalar field for any $D$ for $\a=1$, 
$\left\{\frac{1}{2t}\left[\sqrt{1+4t(X/2)^{D-1}}-1\right]\right\}^{\frac{1}{D-1}}$, but otherwise the deformation can be considered for 
any $\a$ and any potential $V$. 

The subsequent paper of Marika Taylor \cite{Taylor:2018xcy} considers, instead of the two possibilities above, another one, 
a quadratic form similar to the above, but where the trace part has the coefficient $1/(D-1)$, 
\be
\d_t S=\int d^Dx\sqrt{g}\left[T^{\mu\nu}T_{\mu\nu}-\frac{1}{D-1}{T^\mu}_\mu {T^\nu}_\nu\right]\;,
\ee
and argues that it is the correct generalization in view of the holographic proposal of McGough 
et al. \cite{McGough:2016lol}.\footnote{Note that in \cite{Babaei-Aghbolagh:2020kjg} a deformation for which the 
trace part has coefficient $2/D$ was considered as the correct one for deforming a Maxwell theory of Abelian $p$-forms 
to a Born-Infeld-type one, including cases with S-duality, similar to what happens in $D=2$ and $D=4$.}

But we can consider that the same arguments used by  \cite{Araujo:2018rho} to find that the single-trace deformation on the worldsheet,
giving before a $CFT_2$ $T\bar T$ deformation, is equal to a TsT transformation in the boundary directions, can be used to 
argue that the same is true for $AdS_5\times S^5$: the $T\bar T$ deformation is defined by the TsT transformation. 

Furthermore, in the \(AdS_5\times \mathbb{S}^5\) case, the TsT transformation is (conjectured to be) dual to 
noncommutative deformations of the ${\cal N}=4$ Super
Yang-Mills theory~\cite{Hashimoto:1999ut, Maldacena:1999mh, Matsumoto:2014gwa, vanTongeren:2015uha, Araujo:2017jkb}.
In order to obtain a metric that is Lorentz invariant (though, of course, the B field is not), one has to consider 
noncommutativity in the (01) and (23) directions, corresponding to two successive TsT transformations, one in the (01) directions, 
followed by another in the (23) directions. As in \cite{Hashimoto:1999ut, Maldacena:1999mh}, the construction must be 
done in Euclidean signature, so that the results in the $(01)$ and $(23)$ directions are the same, and that there are no 
apparent singularities (in Minkowski signature, the $(01)$ TsT transformation results in a factor of $1-\gamma^2e^{4\rho}$
instead of $1+\gamma^2e^{4\rho}$, even though TsT is a symmetry of the string theory).

For a single TsT transformation, in the (01) Euclidean directions, we start with $B_{ij}=0$, $\Phi=\Phi_0$
and 
\be
ds^2=e^{2\rho}(dt^2+d\vec{x}^2)+d\rho^2+ds^2_{S^5}.
\ee

Then after T-duality on $t$, we still find $B=0$, and 
\bea
ds^2&=& e^{-2\rho}dt^2+e^{2\rho}d\vec{x}^2+d\rho^2+ds^2_{S^5}\cr
\Phi&=&\Phi_0-\frac{1}{2}\log e^{2\rho}.
\eea

After the shift $x\rightarrow x+\gamma t$, where $x=x_1$, we still have $B=0$, and 
\bea
ds^2&=& dt^2(e^{-2\rho}+\gamma^2e^{2\rho})+2\gamma e^{2\rho} dt dx+e^{2\rho}d\vec{x}^2+d\rho^2 +ds^2_{S^5}\cr
\Phi&=&\Phi_0-\frac{1}{2}\log e^{2\rho}.
\eea

Finally, after the T-duality back on $t$, we have 
\bea
ds^2&=& \frac{e^{2\rho}(dt^2+dx^2)}{1+\gamma^2 e^{4\rho}}+e^{2\rho}(dx_2^2+dx_3^2)+d\rho^2+ds^2_{S^5}\cr
B_{01}&=&\frac{\gamma e^{2\rho}}{e^{-2\rho}+\gamma^2 e^{2\rho}}\Rightarrow
H_{01\rho}=\d_\rho B_{01}=-\frac{4\gamma}{(e^{-2\rho}+\gamma^2 e^{2\rho})^2}\cr
\Phi&=& \Phi_0-\frac{1}{2}\log e^{2\rho}-\frac{1}{2}\log (e^{-2\rho}+\gamma^2e^{2\rho})\Rightarrow
e^{2\Phi}=e^{2\Phi_0}\frac{e^{-2\rho}}{e^{-2\rho}+\gamma^2 e^{2\rho}}.
\eea

But rather, for a TsT transformation in the directions (01) and one in the directions (23), 
but with the same parameter, we obtain, like in 
\cite{Hashimoto:1999ut,Maldacena:1999mh}, a Lorentz invariant metric, though with a non-Lorentz invariant B field, 
\bea
ds^2&=& \frac{e^{2\rho}(-dt^2+d\vec{x}^2)}{1+\gamma^2 e^{4\rho}}+d\rho^2+ds^2_{S^5}\cr
B_{01}&=&B_{23}=\frac{\gamma e^{2\rho}}{e^{-2\rho}+\gamma^2 e^{2\rho}}\Rightarrow
H_{23\rho}=H_{01\rho}=\d_\rho B_{01}=-\frac{4\gamma}{(e^{-2\rho}+\gamma^2 e^{2\rho})^2}\cr
\Phi&=& \Phi_0-\log e^{2\rho}-\log (e^{-2\rho}+\gamma^2e^{2\rho})\Rightarrow
e^{2\Phi}=e^{2\Phi_0}\left(\frac{e^{-2\rho}}{e^{-2\rho}+\gamma^2 e^{2\rho}}\right)^2.\label{defads5}
\eea

This is consistent with the results in \cite{Hashimoto:1999ut,Maldacena:1999mh}. We can now Wick rotate back 
to Minkowski signature.

\section{Penrose limit of single-trace $T\bar T$ deformations of $AdS_3\times S^3\times T^4$ and 
$AdS_5\times S^5/{\cal N}=4$ SYM correspondence}

In this section, we want to understand the proposed single-trace $T\bar T$ deformation of $AdS_5\times S^5$ from the 
point of view of ${\cal N}=4$ SYM. When a gauge/gravity duality is difficult to understand, it helps to consider the Penrose 
limit, which will simplify both siders of the duality. Moreover, it is interesting to understand what, if any, is the deformation of the 
spin chain in the BMN limit?

\subsection{Review of Penrose limit for $AdS_5\times S^5$ in Poincar\'{e} coordinates}

Since the metric (\ref{defads5}) is written in Poincar\'{e} coordinates, it is worth first reviewing the method for dealing with 
Penrose limits in Poincar\'{e} coordinates, since it is somewhat unusual. Note that we could consider 
transforming the metric in the equivalent of AdS global coordinates, since then the map to the CFT side 
is better understood. It is not clear if that is equivalent to what we do here.

Unlike the Penrose limit in global coordinates, where the limit is easy to do, and the null geodesic sits in the center of AdS, at $\rho=0$, 
the Penrose limit in AdS in Poincar\'{e} coordinates involves motion in an extra coordinate. The method was developed in 
\cite{PandoZayas:2002dso} and later, in more specific detail, in \cite{Gimon:2002sf}, where it was done for the (near horizon limit of)
D$p$-branes in Poincar\'{e} coordinates, that includes the $AdS_5\times S^5$ case. It was further used in similar cases 
in \cite{Itsios:2017nou}. In both cases, the essential point is that the null geodesic involves, besides $t$, motion in {\em two} spatial 
coordinates, one being the isometry direction and the other the radial direction. 

That is different from the case of motion along a single isometry direction, discussed for instance in \cite{Araujo:2017hvi}.
In this latter case, for motion in $x^\lambda$, $\frac{dx^i}{d\lambda}=\delta^i_\lambda$, so we need no acceleration in that direction, 
leading to the conditions (imposed on the geodesic)
\be
\Gamma^i_{\lambda\lambda}=0.
\ee

For static and diagonal metrics, $\d_t g_{ij}=0$ and $g_{0i}=g^{0i}=0$, the geodesic equation for $i=t$ is always satisfied, and 
for motion in an {\em isometry} direction, $\d_\lambda g_{\mu\nu}=0$, we obtain the simple condition
\be
g^{il}\d^l g_{\lambda\lambda}=\d^i g_{\lambda\lambda}=0\;,\;\;\forall i.
\ee

Anyway, back to our more general case, for motion in two directions, following  \cite{Gimon:2002sf}, for the case $p=3$, i.e., 
for D3-branes giving $AdS_5\times S^5$ in Poincar\'{e} coordinates, with metric 
\be
ds^2=R^2\frac{(-dt^2+d\vec{x}_3^2+dz^2)}{z^2}+R^2d\Omega_5^2\;,\label{AdS5}
\ee
where we write the metric in $S^5$ by separating in an angle, plus an $S^4$, 
\be
d\Omega_5^2=d\psi^2+\sin^2\psi d\Omega_4^2.
\ee

We consider the null geodesic for motion in $(t,z,\psi)$. It is not completely obvious that this will give the same as the motion 
at fixed $\rho=0$ in global coordinates (in the center of AdS), but we find this is true after the fact, since we get the same pp wave.

The method involves: 1) writing an effective Lagrangian for motion in $(t,z,\psi)$. From it, we find the geodesic in terms of a 
null affine parameter $\lambda$, that will be $x^+$ in the end. 2) Then, in terms of that motion on the geodesic, we change 
coordinates from $(t,z,\psi)$ to $(\lambda,\b,\phi)$ defined such that we end up with the metric in the form stated by Penrose 
as always possible, and leading to the pp wave in Rosen coordinates. 3) We take the limit to get the pp wave in Rosen coordinates, 
and then 4) apply the usual transformation to get the Brinkmann coordinates.

Applying it for the metric (\ref{AdS5}), we get: {\bf 1)} the effective Lagrangian for $(t,z,\psi)$ motion (taking out the irrelevant $R^2$ factor)
\be
L=-z^{-2}\dot t^2 +z^{-2}\dot z^2+\dot\psi^2\;,
\ee
which must be taken together with the constraint $L=0$, for a null geodesic ($ds^2=0$). Since $L$ is independent of $t,\psi$
(only on the derivatives), the corresponding equations of motion are integrated to integrals of motion, 
\be
\frac{\d L}{\d \dot t}=-2E={\rm constant}\;,\;\;
\frac{\d L}{\d \dot\psi}=2\mu={\rm constant}\;,
\ee
which together with the constraint $L=0$ give 3 differential equations for $t(\lambda),z(\lambda),\psi(\lambda)$, 
\be
\dot t=z^2 E\;,\;\;
\dot\psi=\mu\;,\;\;
\dot z^2=\dot t^2-z^2\dot\psi^2\Rightarrow \dot z=\pm z\sqrt{z^2 E^2-\mu^2}.
\ee

Then we get 
\bea
\mu \lambda &=&\int^{Ez/\mu}\frac{dy}{y\sqrt{y^2-1}}\;,\cr
\frac{d\psi}{dz}&=&\pm\frac{\mu}{z\sqrt{z^2E^2-\mu^2}}\cr
\frac{dt}{dz}&=&E\frac{z}{\sqrt{z^2E^2-\mu^2}}\;,
\eea
integrated to 
\bea
z&=&z_\lambda\cr
t&=& t_\lambda -\frac{\b}{E}+\mu\phi\cr
\psi&=&\psi_\lambda+E\phi\;,
\eea
where $t_\lambda $ and $z_\lambda$ refer to $\int \frac{dt}{d\lambda}d\lambda$ and $\int \frac{dz}{d\lambda}d\lambda$, respectively, 
and the coefficients of the $\b, \phi$ extra terms (independent of $\lambda$, so "constants" from the point of view of the previous 
integration) are chosen such that this change of coordinates gives $g_{\lambda\b}=1$, $g_{\lambda\phi}=0$, as we will shortly see.

Indeed, by differentiation we obtain 
\bea
dt&=&\frac{dt}{d\lambda}d\lambda -\frac{d\b}{E}+\mu d\phi=z^2Ed\lambda -\frac{d\b }{E}+\mu d\phi\cr
d\psi&=& \frac{d\psi}{d\lambda}d\lambda+Ed\phi=\mu d\lambda+E d\phi\cr
dz&=&\frac{dz}{d\lambda}d \lambda=z\sqrt{z^2E^2-\mu^2}d\lambda\;,
\eea
and {\bf 2)} by substituting this in the metric (\ref{AdS5}), we obtain 
\be
R^{-2}ds^2=2d\lambda d\b-\frac{d\b^2}{E^2z^2}+\frac{2\mu}{E}\frac{d\b d\phi}{z^2}+d\phi^2\left(E^2-\frac{\mu^2}{z^2}\right)+\frac{d\vec{x}^2}
{z^2}+\sin^2(\psi_\lambda+E\phi)d\Omega_4^2\;,
\ee
which is exactly in the form stated by the Penrose theorem ($2dV dU+\a dV^2+\sum_i \b_i dV dY^i+\sum_{i,j}C_{i,j}dY^i dY^j$). 

We can therefore {\bf 3)} take the Penrose limit by rescaling with $R$ and taking $R\rightarrow\infty$, 
\be
U=\lambda =u\;,\;\;
V=\b =\frac{v}{R^2}\;,\;\;
Y^i=\frac{y^i}{R}\;,
\ee
where $Y^i$ stands for $\Omega_4, \phi,\vec{x}_3$. Specifically then 
\be
\vec{\Omega}_4=\frac{\vec{y}_4}{R}\;,\;\;
\phi=\frac{\varphi}{R}\;,\;\;\;
\vec{x}_3=\frac{\vec{x}'_3}{R}.
\ee

The resulting pp wave in Rosen coordinates is 
\be
ds^2=2dudv+d\varphi^2\left(E^2-\frac{\mu^2}{z^2}\right)+\frac{d\vec{x'}_3^2}{z^2}+\sin^2\psi_\lambda d\vec{y}_4^2\;,
\ee
and where $z=z(\lambda=u),\psi_\lambda=\psi_\lambda(\lambda=u)$.

For the {\bf 4)}  transformation to Brinkmann coordinates, we have in general 
\bea
g_{ij}&\equiv& e_i^a e_j^b\delta_{ab}\Rightarrow\cr
u&=& x^+\cr
v&=& x^-+\frac{1}{2}\dot e_{ai}e^i_b x^a x^b\cr
y^i&=& e^i_ax^a\;,
\eea
leading to the Brinkmann form pp wave,
\bea
ds^2&=& 2dx^+dx^-+H(x^+,x^i)(dx^+)^2+d\vec{x}^2\cr
H&=& A_{ab}x^ax^b\cr
A_{ab}&=&\ddot e_{ai} e^i_b.
\eea

In our case, we obtain the vielbeins
\be
e^\varphi_\varphi=\sqrt{E^2-\frac{\mu^2}{z^2}}\;,\;\;
e^{x'}_{x'}=\frac{1}{z}\;,\;\;
e^y_y=\sin\psi_\lambda\;,
\ee
so the new coordinates are
\be
\lambda=u=x^+\;,\;\;
\tilde\varphi=\varphi\sqrt{E^2-\frac{\mu^2}{z^2}}\;,\;\;
\tilde x=\frac{x}{z}\;,\;\;
\tilde y=y\sin\psi_\lambda\;,\;\;
z=z(x^+)\;,
\ee
and $x^-$ is not needed. 

Then 
\be
A_{\tilde y \tilde y}=\frac{1}{\sin\psi_\lambda}\frac{d^2}{d\lambda^2}\sin\psi_\lambda=\frac{1}{\sin\psi_\lambda}\frac{d}{d\lambda}
\left(\cos\psi_\lambda\frac{d\psi}{d\lambda}\right)=\tan^{-1}\psi_\lambda \frac{d^2\psi_\lambda}{d\lambda}-\left(\frac{d\psi_\lambda}
{d\lambda}\right)^2=0-\mu^2\;,
\ee
and given that 
\be
\frac{d}{d\lambda}\sqrt{E^2-\frac{\mu^2}{z^2}}=\frac{\mu^2}{z}\;,
\ee
we obtain also
\bea
A_{\tilde x\tilde x}&=&z\frac{d^2}{dz^2}\frac{1}{z}=-z\frac{d}{d\lambda}\left(\frac{1}{z^2}\frac{dz}{d\lambda}\right)
=-z\frac{d}{d\lambda}\sqrt{E^2-\frac{\mu^2}{z^2}}=-\mu^2\cr
A_{\tilde \varphi\tilde\varphi}&=&\frac{1}{\sqrt{E^2-\frac{\mu^2}{z^2}}}\frac{d^2}{d\lambda^2}\sqrt{E^2-\frac{\mu^2}{z^2}}
=\frac{1}{\sqrt{E^2-\frac{\mu^2}{z^2}}}\frac{d}{d\lambda}\frac{\mu^2}{z}=-\frac{\mu^2}{z\sqrt{E^2-\frac{\mu^2}{z^2}}}\frac{dz}{d \lambda}
=-\mu^2\;,\cr
&&
\eea
so in the end we obtain the usual pp wave,
\be
ds^2=2dx^+dx^- -\mu^2(\tilde{\vec{x}}_3\tilde{\vec{x}}_3+\tilde\varphi\tilde\varphi
+\tilde{\vec{y}}_4\tilde{\vec{y}}_4)(dx^+)^2
+d\tilde{\vec{x}}_3^2+d\tilde\varphi^2+d\tilde{\vec{y}}_4^2.
\ee

\subsection{Penrose limit of single-trace $T\bar T$ deformation of $AdS_3\times S^3\times T^4$}

As a warm-up exercise, we first do the Penrose limit on the single-trace $T\bar T$ deformation of $AdS_3\times S^3\times T^4$, 
where we know for sure that the deformation is given by the TsT transformation. 

The metric is (re-introducing the factor of $R^2$ which was needed for the pp wave limit, as it is taken to infinity, 
now corresponding to $R^2_{\rm AdS}=kl_s^2$, and writing the metric on $S^3$ as above, in terms of a
$\psi$ and an $S^2$)
\be
R^{-2}ds^2=\frac{e^{2\rho}(-dt^2+dx^2)}{1+2\gamma e^{2\rho}}+d\rho^2+\frac{1}{4}\left(d\psi^2+\sin^2\psi d\Omega_2^2\right)
+ds^2(T^4).\label{defAdS3}
\ee

Note that, strictly speaking, $R=R_{\rm AdS}\rightarrow \infty $ can be obtained only for $k\rightarrow \infty$, in which 
case we cannot 
observe any  potential phase transition in $k$. 

Then {\bf 1)} the effective Lagrangian for motion in $(t,\rho,\psi)$ is (ignoring the overall $R^2$)
\be
L=-\frac{e^{2\rho}}{1+2\gamma e^{2\rho}}\dot t^2+\dot \rho^2+\frac{1}{4}\dot\psi^2.
\ee

It is independent (explicitly) on $t$ and $\psi$, so the Lagrange equations of motion are integrated with integrals of motion, 
\be
\frac{\d L}{\d \dot t}=-2\frac{\dot t e^{2\rho}}{1+2\gamma e^{2\rho}}=-2E={\rm constant}\;,\;\;\;
\frac{\d L}{\d \dot \psi}=\frac{1}{2}\dot\psi=2\mu={\rm constant}\;,
\ee
giving 
\be
\dot t=e^{-2\rho}(1+2\gamma e^{2\rho})E\;,\;\; \dot\psi=4\mu.
\ee

The constraint $L=0$ (from $ds^2=0$, null geodesic) gives
\be
\dot \rho^2=\frac{e^{2\rho}}{1+2\gamma e^{2\rho}}\dot t^2-\frac{1}{4}\dot\psi^2\Rightarrow
\dot \rho=\pm e^{-\rho}\sqrt{E^2(1+2\gamma e^{2\rho})-4\mu^2 e^{2\rho}}\;,
\ee
so that we obtain 
\bea
\lambda&=&\pm \int \frac{e^{\rho}d\rho}{\sqrt{E^2(1+2\gamma e^{2\rho})-4\mu^2e^{2\rho}}}=\int \frac{d(e^{\rho})}{\sqrt{E^2-e^{2\rho}(4\mu^2
-2\gamma)}}\cr
&=&\frac{1}{4\mu^2-2\gamma}\arcsin \left(\frac{e^\rho\sqrt{4\mu^2-2\gamma}}{E}\right)\;,\;\; {\rm if}\;\; 4\mu^2-2\gamma>0\cr
\frac{d\psi}{d\rho}&=& \pm 4\mu \frac{e^{\rho}}{\sqrt{E^2(1+2\gamma e^{2\rho})-4\mu^2 e^{2\rho}}}\cr
\frac{dt}{d\rho}&=& E\frac{e^{-\rho}(1+2\gamma e^{2\rho})}{\sqrt{E^2(1+2\gamma e^{2\rho})-4\mu^2e^{2\rho}}}.
\eea

Then the coordinate transformation $(t,\rho,\psi)$ to $(\lambda,\b,\phi)$ that puts $g_{\lambda\b}=1,g_{\lambda\phi}=0$ 
is (we fix the coefficients of $\b,\phi$ in this way)
\bea
d\rho&=& \frac{d\rho}{d\lambda}d\lambda=e^{-\rho}\sqrt{E^2(1+2\gamma e^{2\rho})-4\mu^2 e^{2\rho}} d\lambda\cr
d\psi&=& \frac{d\psi}{d\lambda}d\lambda +Ed\phi=4\mu d\lambda+E d\phi\cr
dt&=& \frac{dt}{d\lambda}d\lambda-\frac{d\b}{E}+\mu d\phi=e^{-2\rho}E(1+2\gamma e^{2\rho})d\lambda-\frac{d\b}{E}+\mu d\phi.
\eea

Indeed, {\bf 2)} substituting the above transformation in the metric (\ref{defAdS3}) we obtain 
\bea
R^{-2}ds^2&=&2d\lambda d\b-\frac{d\b^2}{E^2}\frac{e^{2\rho}}{1+2\gamma e^{2\rho}}+\frac{2\mu }{E}d\b d\phi\frac{e^{2\rho}}{1+
2\gamma e^{2\rho}}\cr
&&+d\phi^2\left(\frac{E^2}{4}-\frac{\mu^2 e^{2\rho}}{1+2\gamma e^{2\rho}}\right)
+\frac{dx^2 e^{2\rho}}{1+2\gamma e^{2\rho}}+\sin^2(\psi_\lambda +E\phi)d\Omega_2^2
+ds^2(T^4).\cr
&&
\eea

Taking the {\bf 3)} Penrose limit as usual by $R\rightarrow\infty$, with 
\be
U=\lambda=u\;,\;\;\;
V=\b=\frac{v}{R^2}\;,\;\; \; Y^i=\frac{y^i}{R}\;,
\ee
where $Y^i$ stands in for $\phi,\Omega_2, x, T^4$, so 
\be
\phi=\frac{\varphi}{R}\;,\;\; \vec{\Omega}_2=\frac{\vec{y}_2}{R}\;,\;\;
x=\frac{x'}{R}\;,\;\;
T_4=\frac{T'_4}{R}\;,
\ee
and where $\rho=\rho(\lambda=u), \psi_\lambda=\psi_\lambda(\lambda=u)$, we obtain the pp wave metric in Rosen coordinates,
\be
ds^2=2dudv +d\varphi^2\left[\frac{E^2}{4}-\frac{\mu^2e^{2\rho(u)}}{1+2\gamma e^{2\rho(u)}}\right]
+\frac{(dx')^2 e^{2\rho(u)}}{1+2\gamma e^{2\rho(u)}} +\sin^2\psi_\lambda(u)d\vec{y}_2^2+ds^2(T'_4).
\ee

The dilaton is unchanged through the coordinate changes, and is 
\be
e^{2\Phi}=\frac{e^{2\Phi_0}}{1-2\gamma e^{2\rho(u)}}.
\ee

The B-field with the correct power of $R$ in front is
\bea
B_{01}dt \wedge dx&=& -R\frac{e^{2\rho(u)}}{1+2\gamma e^{2\rho(u)}} dt\wedge dx\cr
&\rightarrow& R \frac{e^{2\rho(u)}}{1+2\gamma e^{2\rho(u)}}e^{-2\rho(u)}(1+2\gamma e^{2\rho(u)})E du \wedge dx\cr
&=&- E du\wedge dx'
\eea

On the other hand, on the $S^3$, we have 
\be 
H_{S^3}=R^2\frac{1}{4}\sigma_1\wedge \sigma_2\wedge \sigma_3\Rightarrow 
B_{S^3} \sim R^2 \psi d\Omega_1\wedge d\Omega_2\;,
\ee
so in the Penrose limit, we get ($\psi=4\mu x^+$ in the Penrose limit)
\be
B_{S^3}\sim 4\mu x^+\sin ^2 4\mu x^+ dy_1\wedge dy_2\;.
\ee

To go to {\bf 4)} the Brinkmann coordinates, we first define the vielbeins, 
\be
e^\varphi_\varphi=\sqrt{\frac{E^2}{4}-\mu^2\frac{e^{2\rho}}{1+2\gamma e^{2\rho}}}\;,\;\;
e^{x'}_{x'}=\frac{e^\rho}{\sqrt{1+2\gamma e^{2\rho}}}\;,\;\;
e^y_y=\sin\psi_\lambda(u)\;,
\ee
so the coordinate change is 
\bea
\lambda =u&=&x^+\cr
\tilde \varphi&=&\varphi \sqrt{\frac{E^2}{4}-\mu^2\frac{e^{2\rho}}{1+2\gamma e^{2\rho}}}\cr
\tilde x&=& x'\frac{e^\rho}{\sqrt{1+2\gamma e^{2\rho}}}\cr
\tilde y&=& y\sin\psi_\lambda(u)\cr
z&=&z(\lambda=u=x^+)\;,
\eea
and $x^-$ doesn't need to be written. 

Then, for the Brinkmann metric, we find 
\be
A_{\tilde y\tilde y}=\frac{1}{\sin\psi_\lambda}\frac{d^2}{d\lambda^2}\sin\psi_\lambda =
\tan^{-1} \psi_\lambda \frac{d^2\psi_\lambda}{d\lambda^2}-\left(\frac{d\psi_\lambda}{d\lambda}\right)^2=0-16\mu^2.
\ee

Moreover, using that
\be
\frac{d}{d\lambda}\left(\frac{e^\rho}{\sqrt{1+2\gamma e^{2\rho}}}\right)=\frac{\sqrt{E^2(1+2\gamma e^{2\rho})-4\mu^2 e^{2\rho}}}{(1+
2\gamma e^{2\rho})^{3/2}}\;,
\ee
we find 
\bea
A_{\tilde x\tilde x}&=&e^{-\rho}\sqrt{1+2\gamma e^{2\rho}}\frac{d^2}{d\lambda^2}\left(\frac{e^\rho}{\sqrt{1+2\gamma e^{2\rho}}}\right)\cr
&=&e^{-\rho}\sqrt{1+2\gamma e^{2\rho}}\frac{d}{d\lambda}\left[\frac{\sqrt{E^2(1+2\gamma e^{2\rho})-4\mu^2 e^{2\rho}}}{(1+
2\gamma e^{2\rho})^{3/2}}\right]\cr
&=& -\frac{4}{(1+2\gamma e^{2\rho})^2}\left[(1+2\gamma e^{2\rho})(\gamma E^2+\mu^2)-6\gamma \mu^2 e^{2\rho}\right]\;,
\eea
and also
\bea
A_{\tilde \varphi\tilde\varphi}&=&\frac{1}{\sqrt{\frac{E^2}{4}-\mu^2\frac{e^{2\rho}}{1+2\gamma e^{2\rho}}}}\frac{d^2}{d\lambda^2}
\sqrt{\frac{E^2}{4}-\mu^2\frac{e^{2\rho}}{1+2\gamma e^{2\rho}}}\cr
&=&\frac{1}{\sqrt{\frac{E^2}{4}-\mu^2\frac{e^{2\rho}}{1+2\gamma  e^{2\rho}}}}\frac{d}{d\lambda}\left[-4\mu^2\frac{e^\rho}{(1+
2\gamma e^{2\rho})^{3/2}}\right]\cr
&=&-8\mu^2\frac{1-4\gamma e^{2\rho}}{(1+2\gamma e^{2\rho})^2}.
\eea

The pp wave metric is then 
\be
ds^2=2dx^+dx^-+H(x^+)(dx^+)^2+d\tilde\varphi^2+d\tilde x^2+d\tilde{\vec{y}}_2^2+ds^2(T^4)\;,
\ee
and 
\be
H(x^+)=A_{\tilde\varphi\tilde \varphi}\tilde\varphi^2+A_{\tilde x\tilde x}\tilde{x}_2^2+A_{\tilde y \tilde y}\tilde{\vec{y}}^2_2.
\ee

Also, we have found $\lambda(\rho)$, and we saw that $\lambda=x^+$, so inverting it, we get
\be
e^{\rho(x^+)}=\frac{E}{\sqrt{4\mu^2-2\gamma}}\sin \left(x^+\sqrt{4\mu^2-2\gamma}\right)\;,\;\;{\rm if}\;\; 4\mu^2-2\gamma>0.
\ee

The string action is written, as usual, in the gauge $x^+=\tau$, and the conformal (unit) gauge $\sqrt{h}h^{ab}=\eta^{ab}$, 
and with $\epsilon^{01}=+1$, we get 
\bea 
S_{\rm string}&=&-\frac{1}{4\pi \a'}\int_0^{2\pi \a' p^+}d\sigma\int d\tau \left[\eta^{ab}G_{\mu\nu}\d_a X^\mu \d_b X^\nu
+\epsilon^{ab} B_{\mu\nu} \d_a X^\mu \d_b X^\nu+2\pi \a' \Phi {\cal R}^{(2)}\right]\cr
&=& -\frac{1}{4\pi \a'}\int_0^{2\pi \a' p^+}d\sigma \int d\tau \left[\eta^{ab}\sum_{i\neq \pm}\d_a X^i\d_b X^i -8\mu^2\tilde\varphi^2
\frac{1-4\gamma  e^{2\rho(\tau)}}{1+2\gamma e^{2\rho(\tau)}}-16\mu^2\tilde{\vec{y}}^2_2\right.\cr
&&\left.-4\tilde x^2\frac{(1+2\gamma e^{2\rho(\tau)})(\mu^2+\gamma E^2)-6\gamma \mu^2 e^{2\rho}}{(1+2\gamma e^{2\rho(\tau)})^2}
\right.\cr
&&\left.
-E \d_1 x' +4\mu x^+\sin^2(4\mu x^+)(\d_0 y_1 \d_1 y_2-\d_1 y_1\d_0 y_2)\right].
\eea

Note that $x'=x e^{-\rho}\sqrt{1+2\gamma e^{2\rho}}$ and $y=\tilde y/\sin(4 \mu x^+)=\tilde y/\sin (4\mu \tau)$ in the above 
(for the B field).

\subsection{Penrose limit of proposed single-trace $T\bar T$ deformation of $AdS_5\times S^5$}

We move on to the most interesting case, of the proposed single-trace $T\bar T$ deformation of $AdS_5\times S^5$, and we repeat the 
procedure.

For the solution 
\bea
R^{-2}ds^2&=& \frac{e^{2\rho}(-dt^2+d\vec{x}_3^2)}{1+\gamma^2 e^{4\rho}}+d\rho^2+d\psi^2+\sin^2\psi d\Omega_4^2\cr
B_{01}=B_{23}&=&\frac{\gamma e^{2\rho}}{e^{-2\rho}+\gamma^2 e^{2\rho}}\cr
e^{2\Phi}&=&e^{2\Phi_0}\left(\frac{e^{-2\rho}}{e^{-2\rho}+\gamma^2e^{2\rho}}\right)^2\;,
\eea
we consider motion in $(t,\rho,\psi)$, with the effective Lagrangian
\be
R^{-2}L=-\frac{e^{2\rho}\dot t^2}{1+\gamma^2 e^{4\rho}}+\dot \rho^2+\dot\psi^2\;,
\ee
and equations of motion
\bea
\frac{\d L}{\d \dot t}&=&-\frac{2e^{2\rho}\dot t}{1+\gamma^2 e^{4\rho}}=-2E\cr
\frac{\d L}{\d \dot \psi}&=& 2\dot\psi=2\mu={\rm const.}\;,
\eea
so 
\bea
\dot t&=& e^{-2\rho}E(1+\gamma^2 e^{4\rho})\cr
\dot\psi&=&\mu\;,
\eea
plus the condition of null motion, so $L=0$, i.e., 
\be
\dot \rho^2=\frac{e^{2\rho}\dot t^2}{1+\gamma^2 e^{4\rho}}-\dot \psi^2\Rightarrow
\dot \rho =\pm  e^{-\rho}\sqrt{E^2(1+\gamma^2 e^{4\rho})-\mu^2 e^{2\rho}}\;,
\ee
integrated to 
\be
\lambda =\pm \int \frac{e^\rho d\rho}{\sqrt{E^2(1+\gamma^2 e^{4\rho})-\mu^2 e^{2\rho}}}=
\frac{1}{E\gamma}\int \frac{d(e^\rho)}{\sqrt{\frac{1}{\gamma^2}
-\frac{\mu^2}{4E^4\gamma^4}+\left(e^{2\rho}-\frac{\mu^2}{2E^2\gamma^2}\right)^2}}.
\ee

We obtain 
\bea
\frac{d\psi}{d\rho}&=&\pm \mu \frac{e^\rho}{\sqrt{E^2(1+\gamma^2e^{4\rho})-\mu^2 e^{2\rho}}}\cr
\frac{dt}{d\rho}&=& \pm E \frac{e^{-\rho}(1+\gamma^2 e^{4\rho})}{\sqrt{E^2(1+\gamma^2e^{4\rho})-\mu^2 e^{2\rho}}}\;,
\eea
leading to the change of coordinates
\bea
d\rho&=&\frac{d\rho}{d\lambda}d\lambda=d\lambda e^{-\rho}\sqrt{E^2(1+\gamma^2e^{4\rho})-\mu^2 e^{2\rho}}\cr
d\psi&=& \frac{d\psi}{d\lambda}d\lambda +Ed\phi=\mu d\lambda +Ed\phi\Rightarrow\cr
\psi&=& \psi_\lambda +E\phi\cr
dt&=& \frac{dt}{d\lambda}d\lambda -\frac{d\b}{E}+\mu d\phi\cr
&=& e^{-2\rho}E(1+\gamma^2 e^{4\rho})d\lambda -\frac{d\b}{E}+\mu d\phi\;,
\eea
which when substituted in the metric leads to (we obtain $g_{\lambda\phi}=g_{\lambda\lambda}=0$, $g_{\lambda\b}=1$)
\bea
R^{-2}ds^2&=&2d\lambda d\b -\frac{d\b^2}{E^2}\frac{e^{2\rho}}{1+\gamma^2e^{4\rho}}+2d\b d\phi\frac{\mu}{E}\frac{e^{2\rho}}{1
+\gamma^2e^{4\rho}}\cr
&&+d\phi^2\left(E^2-\mu^2\frac{e^{2\rho}}{1+\gamma^2e^{4\rho}}\right).
\eea

Take the Penrose limit,
\be
U=\lambda=u\;,\;\;
V=\b =\frac{v}{R^2}\;,\;\;
Y^i=\frac{y^i}{R}\;,
\ee
so specifically
\be
\phi=\frac{\varphi}{R}\;,\;\;
\vec{\Omega}_4=\frac{\vec{y}_4}{R}\;,\;\;
\vec{x}_3=\frac{\vec{x}'_3}{R}\;,
\ee
and $\rho=\rho(\lambda=u)$, $\psi_\lambda=\psi_\lambda(\lambda=u)$. We obtain the pp wave metric in Rosen coordinates,
\bea
ds^2_{\rm Rosen}&=& 2dudv +d\varphi^2\left[E^2-\mu^2\frac{e^{2\rho(u)}}{1+\gamma^2 e^{4\rho(u)}}\right]
+\sin^2(\psi_\lambda(u))d\vec{y}_4^2\cr
&&+(d\vec{x}'_3)^2\frac{e^{2\rho(u)}}{\left(1+\gamma^2 e^{4\rho(u)}\right)^2}.
\eea

For the B field, we find (since $dt\wedge dx_1+dx_2\wedge dx_2\rightarrow \frac{1}{R}Ee^{-2\rho(u)}
(1-\gamma^2e^{4\rho(u)})du\wedge dx'_1$)
\bea
B&=&R\frac{\gamma }{1+\gamma^2 e^{4\rho}}(dt\wedge dx_1+dx_2\wedge dx_3)\cr
&=&\gamma Ee^{2\rho(u)}du\wedge dx'_1\;,
\eea
and the dilaton remains the same, 
\be
e^{2\Phi}=\frac{e^{2\Phi_0}}{\left(1+\gamma^2e^{4\rho(u)}\right)^2}.
\ee

To go to the Brinkmann coordinates, we note that 
\be
e_\varphi^\varphi=\sqrt{E^2-\mu^2\frac{e^{2\rho}}{1+\gamma^2 e^{4\rho}}}\;,\;\;
e_{x_i'}^{x_j'}=\frac{e^{\rho}}{\sqrt{1+\gamma^2e^{4\rho}}}\delta_i^j\;,\;\;
e_{y_i}^{u_j}=\sin\psi_\lambda(u)\delta_i^j\;,
\ee
so the coordinate transformation is 
\bea
\lambda=u=x^+\;,&&\tilde y=u \sin\psi_\lambda(u)\;,\cr
\tilde\varphi=\varphi\sqrt{E^2-\mu^2\frac{e^{2\rho}}{1+\gamma^2e^{4\rho}}}\;,&&
z=z(\lambda=u=x^+)\;,\cr
\tilde{\vec{x}}=\vec{x}'\frac{e^\rho}{\sqrt{1+\gamma^2e^{4\rho}}}.&&
\eea

Then, we find 
\be
A_{\tilde y\tilde y}=\frac{1}{\sin\psi_\lambda}\frac{d^2}{d\lambda^2}\sin\psi_\lambda=
\tan^{-1}\psi_\lambda\frac{d^2\psi_\lambda}{d\lambda^2}-\left(\frac{d\psi_\lambda}{d\lambda}\right)^2=0-\mu^2.
\ee

Moreover, using that
\be
\frac{d}{d\lambda}\left(\frac{e^\rho}{\sqrt{1+\gamma^2e^{4\rho}}}\right)=
\sqrt{E^2(1+\gamma^2e^{4\rho})-\mu^2e^{2\rho}}\frac{1-\gamma^2e^{4\rho}}{(1+\gamma^2e^{4\rho})^{3/2}}\;,
\ee
we find 
\bea
A_{\tilde x\tilde x}&=&e^{-\rho}\sqrt{1+\gamma^2e^{4\rho}}\frac{d^2}{d\lambda^2}\frac{e^\rho}{\sqrt{1+\gamma^2e^{4\rho}}}\cr
&=&e^{-\rho}\sqrt{1+\gamma^2e^{4\rho}}\frac{d}{d\lambda}
\left(\sqrt{E^2(1+\gamma^2e^{4\rho})-\mu^2e^{2\rho}}\frac{1-\gamma^2e^{4\rho}}{(1+\gamma^2e^{4\rho})^{3/2}}\right)\cr
&=&-\frac{E^2(1+\gamma^2e^{4\rho})-\mu^2 e^{2\rho}}{(1+\gamma^2 e^{4\rho})^2} 2\gamma^2 e^{2\rho}(3-\gamma^2e^{4\rho})
-(-2E^2\gamma^2e^{2\rho}+\mu^2)\frac{1-\gamma^2e^{4\rho}}{1+\gamma^2e^{4\rho}}\;,\cr
&&
\eea
and also
\bea
A_{\varphi\varphi}&=&\frac{1}{\sqrt{E^2-\mu^2\frac{e^{2\rho}}{1+\gamma^2 e^{4\rho}}}}\frac{d^2}{d\lambda^2}\sqrt{E^2-\mu^2
\frac{e^{2\rho}}{1+\gamma^2e^{4\rho}}}\cr
&=&\frac{1}{\sqrt{E^2-\mu^2\frac{e^{2\rho}}{1+\gamma^2 e^{4\rho}}}}\frac{d}{d\lambda}\left[-2\mu^2\frac{e^{\rho}(1-\gamma^2 e
^{4\rho}}{(1+\gamma^2e ^{4\rho})^{3/2}}\right]\cr
&=&-2\mu^2\frac{1-8\gamma^2e^{4\rho}-\gamma^4e^{8\rho}}{(1+\gamma^2e^{4\rho})^2}.
\eea

Then the pp wave metric is 
\be
ds^2=2dx^+dx^-+\left[A_{\varphi\varphi}\varphi^2+A_{\tilde x\tilde x}\vec{\tilde x}_3^2+A_{\tilde y\tilde y}\vec{\tilde y}_4^2\right]
(dx^+)^2+d\tilde\varphi^2+(d\vec{\tilde x}_3)^2+(d\vec{\tilde y}_4)^2\;,
\ee
and the string action, in the light-cone gauge with $x^+=\tau$,  is 
\bea 
S_{\rm string}&=&-\frac{1}{4\pi \a'}\int_0^{2\pi \a' p^+}d\sigma\int d\tau \left[\eta^{ab}G_{\mu\nu}\d_a X^\mu \d_b X^\nu
+\epsilon^{ab} B_{\mu\nu} \d_a X^\mu \d_b X^\nu+2\pi \a' \Phi {\cal R}^{(2)}\right]\cr
&=& -\frac{1}{4\pi \a'}\int_0^{2\pi \a' p^+}d\sigma \int d\tau \left\{\eta^{ab}\sum_{i\neq \pm}\d_a X^i\d_b X^i -2\mu^2\tilde\varphi^2
\frac{1-8\gamma^2e^{4\rho(\tau)}-\gamma^4e^{8\rho(\tau)}}{(1+\gamma^2e^{4\rho(\tau)})^2}
\right.\cr
&&\left.-\mu^2\tilde{\vec{y}}^2_4
-\vec{\tilde x}_3^2\left[\left(-2E^2\gamma^2e^{2\rho(\tau)}+\mu^2\right)
\frac{1-\gamma^2e^{4\rho(\tau)}}{1+\gamma^2e^{4\rho(\tau)}}\right.\right.\cr
&&\left.\left.+\frac{E^2\left(1+\gamma^2e^{4\rho(\tau)}\right)-\mu^2 e^{2\rho(\tau)}}{(1+\gamma^2 e^{4\rho(\tau)})^2} 2\gamma^2 e^{2\rho(\tau)} 
\left(3-\gamma^2e^{4\rho(\tau)}\right)\right]\right\}
+\gamma e^{2\rho(\tau)}\d_\sigma x_1'\;,
\eea
and note that in the above, $x'=\tilde x e^{-\rho}\sqrt{1+\gamma^2e^{4\rho}}$ (for the term coming from the B field). 

As a simple check, note that the $\gamma\rightarrow 0$ limit of the above reduces to the string on the maximally supersymmetric 
pp wave, with parameter $\mu$ (the bosonic part of the action).

\subsection{Interpretation in ${\cal N}=4$ SYM of the  deformation}

{\bf Symmetries}

To understand the $T\bar T$ deformation, and more specifically its Penrose limit, consider first the symmetries of the gravitational 
background, and match them against the symmetries in field theory.

In the case of the pp wave of $AdS_5\times S^5$, the initial symmetry of $PSU(2,2|4)$, with bosonic subgroup $SO(4,2)\times SO(6)$
gets changed in the Penrose limit to the pp wave algebra. In particular, $SO(4,2)$ breaks to $SO(4)_1\times SO(2)_1$ (where $SO(2)_1$ 
corresponds to
$x^+$ translations, and $SO(4)_1$ rotates the $\vec{\tilde x}_3$ and $\varphi$, and $SO(6)$ breaks to another 
$SO(4)_2\times SO(2)_2$, where $SO(2)_2$ corresponds to $x^-$ translations, and $SO(4)_2$ rotates the $\vec{\tilde y}_4$.

Of course, at least as far as the bosonic action goes, $SO(4)_1\times SO(4)_2$ is actually enhanced to $SO(8)$, but that is
irrelevant, since the 5-form $F_5$ (the coupling to fermions) breaks $SO(8)$ back to $SO(4)_1\times SO(4)_2$.

Now, in the $T\bar T$ deformed case, for the metric, 
$x^+=\tau$ translation is not a symmetry anymore, so $SO(2)_1$ is gone, and $SO(4)_1$ is further broken to $SO(3)$, 
that rotates $\vec{x}_3$ only (no $\varphi$). However, the B field breaks further $SO(3)$ to $SO(2)'_1$, rotating only $x_2, x_3$.
So, we have the symmetry $SO(2)'_1\times SO(4)\times SO(2)_2$. But, {\em like in the undeformed case}, there must also be some 
translation-type generators $e_i$ and $e^*_i$, or $a_i$ and $a_i^\dagger$, that supplement the breaking of the rotational 
symmetry during the Penrose limit, giving a total number of generators equal to the one before the limit (since the number of generators
cannot decrease in the Penrose limit). 

Indeed, before the 
Penrose limit we have $SO(2)_1\times SO(2)_2\times SO(6)$, for $x_0,x_1$ and $x_2,x_3$ rotations, and for $S^5$ rotations, 
for a total of 1+1+15=17 bosonic generators. But after the Penrose limit we find only 1+1+6= 8, so we miss 9 generators. 
Since the we have the same $-\mu^2\tilde{\vec{y}}_4^2(dx^+)^2$ term as in the undeformed metric, we have the same 4+4 extra
Killing spinors (see \cite{Figueroa-OFarrill:2001hal,Blau:2001ne}; $\d_i=\d/\d\tilde y^i$)
\bea
\xi_{e_i}&=&-\cos (\mu x^+)\d_i -\mu \sin (\mu x^+)\tilde y^i \d_-\cr
\xi_{e^*_i}&=&-\mu \sin (\mu x^+)\d_i+\mu^2\cos(\mu x^+)\tilde y^i\d_-\;.
\eea

Then we can write (as in \cite{Berenstein:2002jq}) $a_i\sim e_i+e_i^*$ and $M_{ij}=x^i\d_j-x^j\d_i=i(a_i^\dagger a_j
-a_j^\dagger a_i)$, with $[a_i,a_j^\dagger]=\delta_{ij} $ (harmonic oscillators) and $e=-p_-$ commuting with everything. 
We also have
$h=-p_+=\mu \sum_i a_i^\dagger a_i$ and this gives the bosonic algebra. We are still missing one generator, but 
this  is just $M_{23}=x_2\d_3-x_3\d_2$.

Since the $SO(4)_2$ symmetry is maintained, and that corresponded
to a R-symmetry rotation of the 4 positive $J$ fermions, it seems to suggest that the supersymmetry is still ${\cal N}=4$.

{\bf Quantization and eigenvalues vs. SYM anomalous dimensions}

The equations of motion of the modes are:
\bea
&&(-\d_\tau^2+\d_\sigma^2-\mu^2)y^i=0\;,\;\; i=1,...,4\cr
&&\left[-\d_\tau^2+\d_\sigma^2-2\mu^2\frac{1-8\gamma^2e^{4\rho(\tau)}-\gamma^4e^{8\rho(\tau)}}
{(1+\gamma^2e^{4\rho(\tau)})^2}\right]
\varphi=0\cr
&&\left\{-\d_\tau^2+\d_\sigma^2-\left[\left(-2E^2\gamma^2e^{2\rho(\tau)}+\mu^2\right)\frac{1-\gamma^2e^{4\rho(\tau)}}{1
+\gamma^2e^{4\rho(\tau)}}
\right.\right.\cr
&&\left.\left.
+\frac{E^2(1+\gamma^2e^{4\rho(\tau)})-\mu^2 e^{2\rho(\tau)}}{(1+\gamma^2 e^{4\rho(\tau)})^2} 2\gamma^2 e^{2\rho(\tau)} 
(3-\gamma^2e^{4\rho(\tau)})\right]\right\}\tilde x_a=0\;,\;\; a=2,3\cr
&&\left\{-\d_\tau^2+\d_\sigma^2-\left[\left(-2E^2\gamma^2e^{2\rho(\tau)}+\mu^2\right)\frac{1-\gamma^2e^{4\rho(\tau)}}
{1+\gamma^2e^{4\rho(\tau)}}
\right.\right.\cr
&&\left.\left.
+\frac{E^2(1+\gamma^2e^{4\rho(\tau)})-\mu^2 e^{2\rho(\tau)}}{(1+\gamma^2 e^{4\rho(\tau)})^2} 2\gamma^2 e^{2\rho(\tau)} 
(3-\gamma^2e^{4\rho(\tau)})\right]\right\}\tilde x_1\cr
&&+\gamma \d_\sigma\left[e^{2\rho(\tau)}e^{-\rho(\tau)}\sqrt{1+\gamma^2e^{4\rho(\tau)}}\right]=0\cr
&&
\eea

We choose as usual for all the modes
\be
X^i=X^i_0\exp[-i\omega \tau+ik_i \sigma]\;,
\ee
and with the usual rescaling by $p^+$ for the gauge condition, we get for the quantization of the momenta around the $\sigma $
circle
\be
k_{i,n}=\frac{n_i}{\a' p^+}.
\ee

Then the only simple modes are the $\tilde y^i$'s, for which we get the usual
\bea
&&(\omega_y^2-k_i^2-\mu^2)\tilde y^i_0=0\Rightarrow \omega_y^2=\mu^2+k_i^2\Rightarrow \cr
&&\frac{\omega_y}{\mu}=\sqrt{1+\frac{n_i^2}{(\mu\a' p^+)^2}}.
\eea

{\bf Discretization and spin chain}




Consider, as usual, a $Z=X^1+iX^2$ field that is charged under the $J=i\d_\psi$ in the 
gravity dual, and the rest we call $\Phi^i$, $i=1,2,3,4$, and their insertions into $\Tr[Z^J]$ correspond to  string modes.

The undeformed case (${\cal N}=4 $ SYM) has a pp wave with $SO(4)_1\times SO(2)_1\times SO(4)_2\times SO(2)_2$ 
symmetry, with the 1 index corresponding to the ${\cal N}=4 $ SYM directions, so the insertions of $D_iZ=\d_i Z+[A_i,Z]$, 
whereas the 2 index 
corresponding to the transverse scalar directions, so the insertions of the $\Phi^i$'s. 

On the other hand, in the $T\bar T$ deformed case, we have a pp wave with $SO(2)'_1\times SO(4)_2\times SO(2)_2$
symmetry, and again the 2 index corresponds to transverse scalar directions, so insertions of the $\Phi^i$'s, and it is unchanged.
Moreover, the interactions coming from the (transverse part of the) 
pp wave are the same, so the interaction Hamiltonian (the potential) in field theory is unchanged, namely $[\Phi^i,\Phi^j]^2$.

But the $D_iZ$ insertions must change, since now we have only $SO(2)\times SO(2)$ symmetry, 
instead of $SO(4)_1$.
That should mean a {\em dipole} theory change in the kinetic term, breaking Lorentz invariance, (01) and (23) being singled 
out. That happens for the case of noncommutative theories, for instance (which are examples of dipole-like theories). 
Also, $SO(4)_2\simeq SU(2)\times SU(2)$, which means that there is one ${\cal N}=2$ supersymmetry acting on half of the 
fermions (fermions are in some complex representation, so of $SU(2)$), and another ${\cal N}=2$ supersymmetry 
on the other half of 
fermions (since Lorentz symmetry is broken). 

Because the TsT deformation was argued in 
\cite{Hashimoto:1999ut, Maldacena:1999mh, Matsumoto:2014gwa, vanTongeren:2015uha, Araujo:2017jkb}
to be dual to the noncommutative (star product) deformation of the field theory, it is very likely, though the Penrose limit
could involve extra subtleties, that the dipole theory is noncommutative. In fact, in \cite{Guica:2017mtd} the 
(null) dipole deformed theory of \cite{Bergman:2000cw} was  related to some TsT deformation, and the full spin chain 
(in the $SL(2)$ sector, away from the BMN limit) was described\footnote{We thank Fedor Levkovich-Maslyuk
for mentioning his work to us, after the first version of our paper appeared on the arXiv.}. 

In conclusion, we have the $T\bar T$ deformation, followed by Penrose limit gives a deformation of the ${\cal N}=4$ SYM 
in the kinetic term, for instance through the change of the usual product with the star product, giving a noncommutative theory, 
but other possibilities for the dipole theory could also happen.

\section{$T\bar T$ deformation of string worldsheet on $AdS_5\times S^5$ pp wave vs. ${\cal N}=4$ SYM spin chain deformation}

Next, we consider the opposite order: first Penrose limit, then $T\bar T$ deformation. That is, we would like to find the 
$T\bar T$ deformation of the BMN sector of ${\cal N}=4$ SYM. There are several ways that could be defined, however.

\subsection{First try: discretization of two-dimensional $T\bar T$ deformed string worldsheet}

Since the BMN spin chain Hamiltonian, describing ${\cal N}=4$ SYM interactions in the BMN limit, is obtained from a discretization of the 
string Hamiltonian on the pp wave, 
written in terms of $\phi_i=\frac{a_i+a_i^\dagger}{\sqrt{2}}$, just that $a_i$ are Cuntz oscillators at a site,
we can try first to discretize the $T\bar T$ deformed string worldsheet Hamiltonian. 

Then, by $T\bar T$ deforming the string worldsheet Hamiltonian on the pp wave and discretizing, we should, almost by 
definition, obtain the $T\bar T$ deformed spin chain in terms of Cuntz oscillators: we could, in fact, {\em define} the deformed 
spin chain like that. The question then is whether this would be useful, meaning whether the resulting Hamiltonian can be 
derived from SYM, or from a deformed SYM, using the same procedure \cite{Berenstein:2002jq} did for the undeformed case. 

In \cite{Bonelli:2018kik}, a $T\bar T$ deformed Lagrangian density for a 2-dimensional scalar with a potential $V$ was given. 
Specializing for the case of just a mass term, $V=\mu^2X^2/2$, it is 
\be
{\cal L}=-\frac{\sqrt{1+2\lambda (\d_\mu X)^2(1-\lambda \mu^2 X^2/2)}-(1-\lambda \mu^2 X^2)}{2\lambda (1-\lambda \mu^2 X^2/2)}.
\ee

If we consider several scalars $X_i$, we can consider the {\em independent $T\bar T$ deformation} for each of them, and 
sum the corresponding Lagrangians, which is what we will do here. 

Therefore when discretizing, and when considering several  coordinates (scalars) $X^I$, we obtain the $T\bar T$ deformed
Lagrangian
\be
L=\int dx {\cal L}\rightarrow \sum_{I,i} L_i^I\;,
\ee
where, since $(\d_\mu X)^2=-\dot X^2+(X')^2$, and $(X')^2$ discretizes as $(X_i-X_{i+1})^2/a^2$ ($a$ is the length of a step on the chain), 
we have (no sum over $i$ and $I$)
\be
L_i^I=-\frac{\sqrt{1+2\lambda (-(\dot X_i^I)^2+(X_i^I-X_{i+1}^I)^2/a^2)(1-\lambda \mu_I^2(X_i^I)^2/2)}-(1-\lambda \mu_I^2 (X_i^I)^2)}
{2\lambda (1-\lambda \mu_I^2(X_i^I)^2/2)}.
\ee

Expanding in $\lambda$, we get 
\bea
L_i^I&=&\frac{1}{2}\left[(\dot X_i^I)^2-(X_i^I-X_{i+1}^I)^2/a^2-\mu_I^2(X_i^I)^2\right]\cr
&&+\lambda \left[\frac{1}{4}\left( (\dot X_i^I)^2-(X_i^I-X_{i+1}^I)^2/a^2\right)^2-\frac{1}{4}\mu_I^4(X_i^I)^4\right]+...
\eea

The canonical momentum is, in the $\lambda$ expansion,
\be
p_i^I=\frac{\d L_i^I}{\d \dot X_i^I}=\dot X_i^I+\lambda \dot X_i^I\left( (\dot X_i^I)^2-(X_i^I-X_{i+1}^I)^2/a^2\right)+...\;,
\ee
and the Hamiltonian, in the same expansion, 
\bea
H_i^I&=& p_i^I\dot X_i^I-L_i^I=\frac{1}{2}[(\dot X_i^I)^2+(X_i^I-X_{i+1}^I)^2/a^2+\mu_I^2 (X_i^I)^2]\cr
&&+\frac{\lambda}{4}\left[3(\dot X_i^I)^4-2(\dot X_i^I)^2(X_i^I-X_{i+1}^I)^2/a^2-(X_i^I-X_{i+1}^I)^4/a^4
\mu_I^4 (X_i^I)^4\right]+...
\eea

Next, in order to find the spin chain Hamiltonian (in terms of Cuntz oscillators at a site $a_i^I$), we need to write 
\be
X_i^I=\frac{a_i^I e^{-i\mu_I t}+(a_i^I)^\dagger e^{i\mu t}}{\sqrt{2}}\;,
\ee
and {\em only at the end of the calculation, after taking the time derivatives}, put $t=0$. In the leading term, we obtain 
\be
H_i^I=\frac{a_i^I (a_i^I)^\dagger +(a_i^I)^\dagger a_i^I}{2}+\frac{1}{a^2}\left(\frac{a_i^I+(a_i^I)^\dagger}{\sqrt{2}}
-\frac{a_{i+1}^I+(a_{i+1}^I)^\dagger}{\sqrt{2}}\right)^2\;,\label{H0i}
\ee
(this was what was obtained in \cite{Berenstein:2002jq}) but the next term looks more complicated. 

The total Hamiltonian $H$ needs to be diagonalized in order to find the spectrum. 
For that, we must define (at least in the leading term, not clear if 
we need to modify this for the other terms) first a Fourier transform, 
\be
a_j =\frac{1}{\sqrt{L}}\sum_{n=1}^L e^{\frac{2\pi ijn}{L}}b_n\;,
\ee
then the mixing of forward and backward (left and right) waves, 
\be
b_n=\frac{c_{n,1}+c_{n,2}}{\sqrt{2}}\;,\;\;
b_{L-n}=\frac{c_{n,1}-c_{n,2}}{\sqrt{2}}\;,
\ee
for $n\leq L/2$, and finally a Bogoliubov transformation mixing $c$'s and $c^\dagger$'s, 
\be
d_{n,i}=\a_{n,i} c_{n,i}+\b_{n,i}c^\dagger_{n,i}\;, i=1,2\;,
\ee
to obtain a diagonal Hamiltonian. Moreover, one can  check that, {\em in the dilute gas approximation}
$d_{n,i}$ {\em approximately} satisfy the usual commutation relations ($[d,d^\dagger]=1, [d,d]=0, [d^\dagger, d^\dagger]=0$), 
and not anymore the Cuntz ones. 

We should check that/if the transformation above still holds, and the commutation relations in the dilute gas approximation 
still hold. 

However, it doesn't hold, as we now show. 

The Hamiltonian is found as follows. For the one-scalar Lagrangian, we find 
\be
{\cal H}(\lambda)=p\dot \phi -{\cal L}=\frac{1}{2\bar\lambda}\frac{1+2\bar\lambda \phi'^2}{\sqrt{1+2\bar\lambda (-\dot\phi^2+\phi'^2)}}
+\tilde V\;,
\ee
where 
\be
\bar\lambda=\lambda(1-\lambda V)\;,\;\;
\tilde V=-\frac{1}{2\bar\lambda}(1-2\lambda V).
\ee

Then, in our case, with $X_i^I$, we have 
\be
p_i^I=\frac{\d L_i^I}{\d \dot X_i^I}=\frac{4\dot X_i^I}{\sqrt{1+2\lambda\left(1-\lambda \frac{\mu^2 (X_i^I)^2}{2}\right)(-(\dot X_i^I)^2+({X'_i}^I)^2)}}
\;,
\ee
and so the Hamiltonian is 
\be
{\cal H}=\frac{1}{2\lambda \left(1-\frac{\lambda\mu^2(X_i^I)^2}{2}\right)}\frac{1+2\lambda\left(1-\frac{\lambda \mu^2(X_i^I)^2}{2}\right)
({X'_i}^I)^2}{\sqrt{1+2\lambda\left(1-\lambda \frac{\mu^2 (X_i^I)^2}{2}\right)(-(\dot X_i^I)^2+({X'_i}^I)^2)}}
-\frac{1-\lambda \mu^2 (X_i^I)^2}{2\lambda\left(1-\frac{\lambda \mu^2(X_i^I)^2}{2}\right)}.
\ee

In order to get the spin chain Hamiltonian, we substitute in it the fields in terms of creation and annihilation operators, so 
\bea
(\dot X_i^I)^2(t=0)&=&\frac{\mu^2}{2}\left[-(a_i^I)^2-(a^{\dagger I}_i)^2+a_i^I a^{\dagger I}_i+a_i^{\dagger I} a_i^I\right]\cr
({X'_i}^I)^2&=&(X_i^I-X_{i+1}^I)^2/a^2\cr
\mu^2 (X_i^I)^2&=& \frac{\mu^2}{2}\left[(a_i^I)^2+(a^{\dagger I}_i)^2+a_i^I a^{\dagger I}_i+a_i^{\dagger I} a_i^I\right].
\eea

The explicit formula is somewhat long, so we don't write it here.

Then, in terms of the Fourier modes of the $a_j^I$, namely $a_j^I =\frac{1}{\sqrt{L}}\sum_{n=1}^L e^{\frac{2\pi ijn}{L}}b_n^I$, 
where if (approximately, for Cuntz oscillators at a site acting on states in the dilute gas approximation) $[a_k,a^\dagger_j]=\delta_{jk}$, 
we obtain $[b_n,b^\dagger_m]=\delta_{nm}$, and 
\be
\sum_{j=0}^{L-1} a^{\dagger I}_j a^I_j=\sum_{n=0}^{L-1} b^{\dagger I}_n b^I_n\;,
\ee
and 
\bea
A_j^I&\equiv & \left[(a_j^I+a_j^{\dagger I})-(a_{j+1}^I+a^{\dagger I}_{j+1})\right]\cr
&=& \sum_{n=0}^{L-1}\frac{1}{\sqrt{L}}\left[e^{\frac{2\pi i nj}{L}}\left(e^{\frac{2\pi i n }{L}}-1\right)b_n^I+
e^{-\frac{2\pi i nj}{L}}\left(e^{-\frac{2\pi i n}{L}}-1\right)b^{\dagger I}_n\right].
\eea

Further, defining the forward and backward (left and right)
waves, $b_n^I=\frac{c_{n,1}^I+c_{n,2}^I}{\sqrt{2}}$, 
$b^I_{L-n}=\frac{c^I_{n,1}-c^I_{n,2}}{\sqrt{2}}$, the commutation relations are once again respected, and moreover
\be
b^{\dagger I}_n  b_n^I +b^{\dagger I}_{L-n}b^I_{L-n}=c^{\dagger I}_{n,1}c^I_{n,1}+c^{\dagger I}_{n,2}c^I_{n,2}\;,
\ee
and 
\bea
A_j^I&=& \sum_{n=0}^{L/2}\frac{1}{\sqrt{L}}\left[e^{\frac{2\pi in j}{L}}\left(e^{\frac{2\pi in}{L}}-1\right)b^I_n+
e^{-\frac{2\pi i nj}{L}}\left(e^{-\frac{2\pi in }{L}}-1\right)b^{\dagger I}_n\right.\cr
&&\left.+e^{-\frac{2\pi in j}{L}}\left(e^{-\frac{2\pi in}{L}}-1\right)b_{L-n}^I
+e^{\frac{2\pi in j}{L}}\left(e^{\frac{2\pi in }{L}}-1\right)b^{\dagger I}_{L-n}\right].
\eea

Until now, the steps are useful, and continue to work in the same way in our Lagrangian.

However, the essential next step doesn't, since it relies on making the sum  $\sum_j (A_j^I)^2$, and in our case, $(A_j)^2$ appears inside 
a complicated expression with square root, and only then it is summed over. 

With 
\bea
\sum_{j=0}^{L_1}(A_j^I)^2&=&\sum_{n=0}^{L/2}\left\{-\left(1-\cos\frac{2\pi n}{L}\right) \left(c^{\dagger I}_{n,1}c^I_{n,1}
+c_{n,1}^I c_{n,1}^{\dagger I}+c_{n,2}^{\dagger I}c_{n,2}^I+c_{n,2}^I c^{\dagger I}_{n,2}\right)\right.\cr
&&\left.
+2\left(1-\cos\frac{2\pi n}{L}\right)\left[(c_{n,1}^I+c_{n,1}^{\dagger I})^2-(c_{n,2}^I-c_{n,2}^{\dagger I})^2\right]\right\}\;,
\eea
we would have the Hamiltonian finally in a form in which we could use a Bogoliubov transformation. 

Indeed, for a general Hamiltonian
\be
H=\b \frac{a a^\dagger +a^\dagger a}{2}\pm \a \frac{(a\pm a^\dagger)^2}{2}=\b\left[\left(1+\frac{\a}{\b}\right)\frac{aa^\dagger+a^\dagger a}{2}
\pm \frac{\a}{2\b}(a^2+(a^\dagger)^2)\right]\;,
\ee
the Bogoliubov transformation is
\be
b=\tilde \a a+\tilde \b a^\dagger\;,
\ee
and if we impose that $[b,b^\dagger]=1$ (like $[a,a^\dagger]=1$), we get $|\a|^2-|\b|^2=1$. Imposing diagonalization, so no $b^2$
or $(b^\dagger)^2$ terms in $H$, we obtain the condition 
\be
\left(1+\frac{\a}{\b}\right)\tilde \a^+\tilde b^*=\pm \frac{\a}{\b}[(\tilde \a^*)^2+(\tilde \b^*)^2].
\ee

If we have $\a,\b\in \mathbb{R}$, we can define $\tilde \a-\tilde \b\equiv \frac{1}{\sqrt{\omega}}$, $\tilde \a+\tilde \b=\sqrt{\omega}$, 
and obtain 
\be
\omega_1=\sqrt{\frac{1-\a/\b}{1+3\a/\b}}\;,\;\;
\omega_2=\sqrt{\frac{1+3\a/\b}{1-\a/\b}}\;,
\ee
and for both values of $\omega$ the diagonal Hamiltonian 
\be
H=\omega \frac{bb^\dagger+b^\dagger b}{2}.
\ee

\subsection{$T\bar T$ deformation of quantum mechanical spin chain in ${\cal N}=4$ SYM }

If the first try was a deformation of the 2-dimensional system, followed by a discretization, 
and it wasn't very successful (very 
useful), in that we couldn't diagonalize the Hamiltonian to find the spectrum, 
we can next try to deform directly the one-dimensional spin chain.

One approach is defined specifically for spin chains. 
In \cite{Pozsgay:2019ekd} and in \cite{Marchetto:2019yyt} it was argued that the previous integrability-preserving deformation
of spin chains defined by Bargheer, Beisert and Loebbert in \cite{Bargheer:2008jt} and \cite{Bargheer:2009xy} is actually a 
$T\bar T$ deformation. 

They claim that first, the seminal paper \cite{Cavaglia:2016oda} defining explicit 
$T\bar T$ deformations  of 2 dimensional QFTs already 
defines them via a Bethe ansatz, so it is worth following the same way to define the $T\bar T$ deformations of spin chains
\cite{Marchetto:2019yyt}.

They say that the deformations of the Bethe-Yang equations 
\be
e^{ip_j R}\prod_{k\neq j}^NS(p_k,p_j)=1\;,
\ee
are via a deformation of the CDD factor appearing in the S-matrix. Specifically, one can keep the S-matrix fixed and 
modify the equations as
\be
e^{ip_j R+i\a(X_j Y-Y_j X)}\prod_{k\neq j}^N S(p_j,p_k)=1\;,
\ee
or equivalently, modify the S-matrix my multiplication with a phase (CDD factor),
\be
S(p_j,p_k)\rightarrow e^{i\a(X_kY_k-X_k Y_j)}S(p_j,p_k).
\ee
These were the integrable deformations of spin chains found by \cite{Bargheer:2008jt} and \cite{Bargheer:2009xy}.

In the case of the $T\bar T$ deformation, the claim is that one can take has $X_j=p_j$ and $Y_j=H(p_j)$, but the arguments
are somewhat indirect (in particular, the fact that integrability must be preserved). 

One should also mention the works \cite{Baggio:2018gct,Sfondrini:2019smd,Frolov:2019nrr} where the $T\bar T$ deformation 
of the {\em worldsheet} string in gravity duals, in particular on pp waves, is considered through an analysis of the 
worldsheet Hamiltonian arising in uniform lightcone gauge, and is found to equal to the TsT transformation in time $t$ 
and one compact direction $\phi$ in {\em spacetime}. For the $AdS_3\times S^3$ case \cite{Baggio:2018gct}, 
in the massless case
one indeed finds the usual \cite{Cavaglia:2016oda,Bonelli:2018kik}
2 dimensional deformation of massless bosons (on the worldsheet), though in the 
massive case and in the $AdS_5\times S^5$ case \cite{Sfondrini:2019smd,Frolov:2019nrr} one finds a difference, 
which is hard to understand physically. We do not understand this method and the discrepancy well enough to 
comment intelligently on it. Instead, in the method used below, such discrepancy is not possible by construction.

However, there is a parallel line of inquiry, one that is also cited by \cite{Marchetto:2019yyt} as previous work,
by David Gross et al. \cite{Gross:2019ach}, \cite{Gross:2019uxi}, in which the $T\bar T$ 
deformation of quantum mechanics is proposed, based on the holographic proposal of McGough et al. \cite{McGough:2016lol}
and $AdS_2$, and a dimensional reduction from $AdS_3$, and a corresponding one on the boundary. 
\footnote{Note that there is also a third deformation, in terms of a bilinear operator, in the papers of Cardy and Doyon \cite{Cardy:2020olv}, 
and Yunfeng Jiang \cite{Jiang:2020nnb}, though it was not developed further, so we will not describe it.}\footnote{See 
also \cite{He:2021dhr} for an alternative derivation of this proposal.}
This is the approach we will follow here.
Note that there is some controversy about the 
applicability of this result to the case with a potential: in \cite{Ebert:2022xfh} it was argued that the 
original 2 dimensional theory must be conformal invariant, so no potential can be present.  
But the controversy is mostly about semantics: In the Gross et al. prescription, the {\em definition} of the 1 dimensional 
analog of $T\bar T$ deformation was such that it coincides with the holographic $AdS_2$ one of \cite{McGough:2016lol}. 
Otherwise, the reduction prescription from 2 dimensions might not work for nonconformal seed theories, as 
 \cite{Ebert:2022xfh} argued.
We will, however, continue applying the original procedure as it was developed.

In sections 3 and 4 of \cite{Gross:2019ach}, the deformation of quantum mechanics is defined as 
\footnote{Note some signs are different with respect to \cite{Gross:2019ach}.
It was easy to check that the signs were wrong, since the $\lambda\rightarrow 0$ doesn't work.
Instead, \cite{Gross:2019ach} have the formula
\be
L=\frac{1-\sqrt{(1-4\lambda\sum_i \dot q_i^2)(1-8\lambda V)}}{4\lambda}\;,\label{theabove}
\ee
which would correspond to taking $\lambda\rightarrow -\lambda$ AND $V\rightarrow -V$ in (\ref{Lagnew}). 
However, in the Hamiltonian, this was not what was considered, so their formula has incorrect signs.

One observation is that the {\em 2 dimensional} $T\bar T$ deformed Lagrangian density is 
\be
{\cal L}=\frac{1-2\lambda V}{2\lambda(1-\lambda V)}-\frac{\sqrt{[1+2\lambda \sum_i(-\dot q_i^2+q_i'^2)]
(1-\lambda V)}}{2\lambda(1-\lambda V)}\;,
\ee
and if we naively put $q'=0$ in (\ref{theabove}), we get a similar (but not quite! the formula above in 2 dimensions 
has the correct sign for $V$ in the 
$\lambda \rightarrow 0 $ limit; it only matches with their Lagrangian for $V=0$, and rescaling $\lambda $ by 2) 
formula to the above, which may be the reason for the confusion. }
\be
H=\frac{1-\sqrt{1-8\lambda\left(\sum_i \frac{p_i^2}{2}+V(\left\{ q_i\right\})\right)}}{4\lambda}\;,
\ee
which leads to
\be
L=-\frac{1-\sqrt{(1+4\lambda\sum_i \dot q_i^2)(1-8\lambda V)}}{4\lambda}\;,\label{Lagnew}
\ee
via
\be
p_i=\dot q_i\sqrt{\frac{1-8\lambda V}{1+4\lambda \sum_i \dot q_i^2}}.
\ee

Further, in \cite{Gross:2019uxi} it was shown that this $T\bar T$ deformation of the quantum mechanics replaces the Hamiltonian $H$
with a function of it, $f(H)$, which means that the eigenfunctions don't change.

Note that, in general, if 
\be
H(q_i,p_i)=f(H_0(q_i,p_i))\;,
\ee
and $H_0$ has conserved quantities, $\dot p_i=0$, so $\frac{\d H_0}{\d q_i}=0$, then also
\be
\frac{\d H}{\d q_i}=f'(H_0)\frac{\d H_0}{\d q_i}=0\;,
\ee
so all classical integrals of motion remain integrals of motion, and therefore a classically integrable system remains integrable 
after the ($T\bar T$, in this case) deformation. 

We see that the Hamiltonian deformation (\ref{defHam}) is very easy to work with. Anything that we did with $H_0$ and $E_0$ 
we can do with $H(\lambda)$ and $E(\lambda)$. In particular, we can write the explicit form in $a,a^\dagger$ oscillators, 
and diagonalize it. 

So the discretized pp wave Hamiltonian (dual to the original spin chain in the dilute gas approximation) (\ref{H0i}) can be 
$T\bar T$ deformed, 
\be
H(\lambda)=\mu\frac{1}{4\lambda\mu}\left[1-\sqrt{1-8\lambda\mu\sum_{i,I}\left(\frac{a_i^I (a_i^I)^\dagger 
+(a_i^I)^\dagger a_i^I}{2}+\frac{1}{a^2}\left(\frac{a_i^I+(a_i^I)^\dagger}{\sqrt{2}}
-\frac{a_{i+1}^I+(a_{i+1}^I)^\dagger}{\sqrt{2}}\right)^2\right)}\right]\;,
\ee
and the deformation can be diagonalized, obtaining (in the $AdS_5\times S^5$ case)
\be
E(\lambda, g^2N, n/J)=\mu\frac{1}{4\lambda\mu}\left(1-\sqrt{1-8\lambda\mu \sqrt{1+\frac{g^2N}{\pi^2}\sin^2\frac{\pi n}{J^2}}}\right).
\ee

Note that the discretization and $T\bar T$ deformation are not commutative. In the previous subsection
we $T\bar T$ deformed first, and then discretized, now we do the opposite. This is besides the noncommutativity of the
$T\bar T$ deformation and Penrose limit, which is also present: here we take the Penrose limit first, then deform, whereas
in the previous section, we first deformed, then took the Penrose limit.

\subsection{Deformation of ${\cal N}=4$ SYM}

We now try to interpret the quantum mechanical $T\bar T$ deformation from the point of view of ${\cal N}=4$ SYM.

{\bf Symmetries and symmetry algebra}

We start with an analysis of the symmetries and their algebra.

Since $H(\lambda)=f(H_0)$, the global symmetries of the Hamiltonian continue to be symmetries of the deformed one. 
In particular, the $SO(2)_1\times SO(4)_1\times SO(2)_2\times SO(4)_2$ with, in the bosonic case, $SO(4)_1\times SO(4)_2$ 
extended to $SO(8)$, continues to hold for the deformed pp wave Hamiltonian, as it was for the usual pp wave Hamiltonian. 
Also true for the ${\cal N}=4$ supersymmetry (or, more precisely, to the 16 supercharges). 

This means that in this case we are dealing with a {\em deformed sector within ${\cal N}=4$ SYM}.

The bosonic symmetry algebra {\em of the string pp wave Hamiltonian on $AdS_5\times S^5$}, matching the one of ${\cal N}=4$ 
SYM, is defined in terms of Killing spinors (see \cite{Blau:2001ne}) for the  pp wave variables (with respect to that paper, we have 
defined $\mu=2\lambda$),
\bea
&&h=\xi_{e^+}=-\d_+\;,\;\;
\xi_{e^-}=-\d_-\;,\cr
&&\xi_{e_i}=-\cos (\mu x^+)\d_i -\mu \sin (\mu x^+)\tilde y^i \d_-\;,\;\; i=1,...,8\cr
&&\xi_{e^*_i}=-\mu \sin (\mu x^+)\d_i+\mu^2\cos(\mu x^+)\tilde y^i\d_-\;,\cr
&&\xi_{M_{ij}}=x_i\d_j-x_j\d_i\;,\;\; i,j=1,...,4 or 5,...,8.
\eea

The algebra is obtained by defining harmonic oscillators $a_i=(e_i+ie_i^*)/\sqrt{2}$, so $[a_i,a_j^\dagger]=\delta_{ij}$, 
then $M_{ij}=i(a_i^\dagger a_j-a_j^\dagger a_i)$ and $H=H_0=-p_+=\mu \sum_i a_i^\dagger a_i$, while $e=-p_-$ commutes 
with everything. 

More precisely, we want to obtain the commutation relations 
\bea
[a_i,a^\dagger_j]&=&e\delta_{ij}\;,\cr
[ih,a_i^\dagger]&=&\mu a_i^\dagger\;,\;\;\;
[-ih,a_i^\dagger]=-\mu a_i\;,\cr
[M_{ij},a_k]&=&-\delta_{ik}a_j+\delta_{jk}a_i\;,\;\;\;
[M_{ij},a^\dagger_k]=-\delta_{ik}a_j^\dagger+\delta_{jk}a_i^\dagger.
\eea

But the symmetry algebra of the $\xi$'s (which we will now call just $e_i, e_i^*, M_{ij}, h$) is, {\em redefining $e_i^*\rightarrow
\mu e_i^*$}, 
\bea
[e_i,e_j^*]&=&(\mu e) \delta_{ij}\cr
[h,e_i]&=& \mu e_i^*\;,\;\; [h, e_i^*]=-\mu e_i\;,\cr
[M_{ij},e_k]&=& -\delta_{ik} e_j+\delta_{jk}e_i\;,\;\;\;
[M_{ij},e_k^*]=-\delta_{ik}e_j^*+\delta_{jk}e_i^*.
\eea

Then, defining 
\be
a_i=a e_i+ib e^*_i\;,\;\;\; a_i^\dagger =a^* e_i^*-ib^*e_i\;,
\ee
we obtain 
\be
[a_i,a_j^\dagger]=(|a|^2+|b|^2)(\mu e)\delta_{ij}\;,
\ee
so 
\be
|a|^2+|b|^2=\frac{1}{\mu} .
\ee

Moreover, 
\bea
[-ih,a_i]&=&-\mu(be_i+ia e_i^*)\cr
&\equiv& -\mu(ae_i+iae_i^*)=-\mu a_i\Rightarrow a=b\cr
[ih,a_i^\dagger]&=&+\mu (b^*e_i^*-ia^*e_i)\cr
&\equiv& +\mu (a^*e_i^*-ib^*e_i)=+\mu a_i^\dagger\Rightarrow a=b.
\eea

So we finally have 
\be
a=b=\frac{1}{\sqrt{2\mu}}\Rightarrow a_i=\frac{e_i+ie_i^*}{\sqrt{2\mu}}\;,\;\;\;
a_i^\dagger=\frac{e_i^*-ie_i}{\sqrt{2\mu}}.
\ee

Then we can represent 
\be
M_{ij}=i(a_i^\dagger a_j-a_j^\dagger a_i)\;,
\ee
and
\be
ih=\frac{\mu}{e}\sum_i a_i^\dagger a_i.
\ee

We can redefine $a_i=\sqrt{e} \tilde a_i$ (since we can treat $e$ as a number, since it is a central charge: it commutes
with everything, it is $=-p_-=-p^+$), and then we have (we write $H$ instead of $h$, to underline the fact that we now talk about the 
$T\bar T$ deformed one dimensional Hamiltonian)
\bea
(\pm i)H&=&\mu \sum_i\tilde a_i^\dagger \tilde a_i\cr
&=&\sum_{n\geq 1} c_n \frac{1}{\lambda}(\lambda H_0)^n\equiv \frac{1}{4\lambda}\left(1-\sqrt{1-8\lambda H_0}\right)\cr
&=& \sum_{n\geq 1} c_n \frac{1}{\lambda}\left[\lambda \mu_0 \sum_i \tilde a_{0,i}^\dagger \tilde a_{0,i}\right]^n\cr
&=&\mu_0 \sum_{n\geq 1}(\lambda \mu_0 )^{n-1}\left[\sum_i \tilde a_{0,i}^\dagger \tilde a_{0,i}\right]^n.\label{HH0a0}
\eea

Here $e_0,\tilde a_{0,i},\tilde a^\dagger_{0,i}$ are the undeformed symmetry generators 
(in some abstract sense, equal to $e, \tilde a_i,\tilde a_i^\dagger$, namely they {\em satisfy the same algebra, though one with $\mu$ 
and another with $\mu_0$}), 
namely their expression in terms of fields are for the undeformed sector. 
The algebra of deformed generators {\em must be the same as of the undeformed generators}. 
The above relation must be solved for $\tilde a_i$ 
in terms of $\tilde a_{0,i}$. One obvious solution is (there doesn't seem to be a solution involving only $a$'s, no $a^\dagger$'s, 
since the noncommutation of the two makes it unlikely to disentangle)
\be
\tilde a_i=\sum_{n\geq 0} \tilde c_n \left[\lambda \mu \sum_j\tilde a_{0,j}^\dagger\tilde a_{0,j}\right]^n\tilde a_{0,i}\;,\label{aa0}
\ee
which follows from (the following defines the constants $c_n$ and $\tilde c_n$)
\be
\frac{1}{4\lambda}\left(1-\sqrt{1-8\lambda x}\right)\equiv \sum_{n\geq 1} c_n \lambda ^{n-1} x^n\Rightarrow
\sqrt{\frac{1}{4\lambda}\left(1-\sqrt{1-8\lambda x}\right)}\equiv \sqrt{x}\sum_{n\geq 0} \tilde c_n \lambda ^n x^n.
\ee

Indeed, we can check that then the deformation of the Hamiltonian is obtained, if we consider that
\be
e=e_0
\ee
(the central charge is undeformed, and it refers to the same $p^+$), 
but $\mu$ is identified with the energy, so $\mu=E(\lambda)$ and $\mu_0=E_0$.

Then the fact that the deformed and undeformed symmetry algebras are the same 
is obtained from the fact that both are written in the same way in terms of the creation and annihilation operators
(we just have the relation (\ref{aa0}) between the $a$'s and $a_0$'s).

Thus we have (\ref{HH0a0}) relating $H$ and $H_0$, (\ref{aa0}) relating $a_i$ with $a_{0,i}$ (and its dagger for $a_i^\dagger$
with $a_{i,0}^\dagger$), $e=e_0$, and 
\be
M_{ij}=\sum_{n\geq 0}\left[\lambda \mu_0 \sum_j \tilde a^\dagger_{0,j}\tilde a_{0,j}\right]^nM_{0,ij}.
\ee

The relation between $a$'s and $a_0$'s
must be one of {\em equivalence} of the two creation/annihilation operator sets, and consequently of their 
Fock spaces.

This symmetry algebra must also be obtained from ${\cal N}=4$ SYM, on the large $J$ charge sector, though 
it wasn't 
written explicitly in terms of fields in  \cite{Berenstein:2002jq}, so we must first do that. 
Next we must define what are the {\em deformed } generators in terms of the ${\cal N}=4$ SYM fields.

When acting on (undeformed) BMN operators (states in the BMN sector), we can represent 
\be
{(\tilde a_{0,i})^\a}_\b=\frac{\delta}{\delta {(\Phi_i)^\a}_\b}\;,\;\;\; {(\tilde a_{0,i}^\dagger)^\a}_\b={(\Phi_i)^\a}_\b{\bf In}\;,
\ee
where $\Phi_i$, $i=1,...,4$ are the 4 scalars that are inserted inside $\Tr [Z^J]$, with their matrix indices, and ${\bf In}$ refers to 
insertion of the matrix element inside the trace.

Next we must define the set of operators of the deformed subsector. 
The vacua must be the same, since moreover $e=e_0$ and this corresponds to having an undeformed $p^+$, so undeformed $J$.
Thus consider the same vacuum $\Tr[Z^J]$, and insert 
$\sum_{n\geq 0} \tilde c_n (\lambda \mu_0 
\sum_j\Phi_j \frac{\delta}{\delta \Phi_j})^n\Phi_i$, and $\tilde c_n$ solved from $c_n$ as suggested above, so 
\be
a^{\dagger i}_m|0\rangle\sim \sum_l \Tr\left[Z^l\left(\sum_{n\geq 0} \tilde c_n \left[\lambda\mu_0 \sum_j \Phi^j \frac{\delta}{\delta
\Phi_j}\right]^n\right)\Phi_iZ^{J-l}\right] e^{\frac{2\pi m l}{J}}.
\ee

In this way, we find that, at the same time 
we have a {\em deformed } BMN sector, and it is {\em equivalent to the original one}. 
Thus $H(\lambda)$ has both the same eigenstates (the undeformed BMN sector), and the new eigenstates (the deformed BMN
sector), since they are supposed to be equivalent. Then $H(\lambda)$ in the undeformed eigenstates gives $E(\lambda)$, whereas 
$H(\lambda)$ in the deformed eigenstates gives $E_0$.

{\bf Comments on the properties of $H_\lambda=f(\lambda, H_0)$}

As was described in detail in \cite{Gross:2019uxi}, once we have $H_\lambda=f(\lambda, H_0)$, with the extra condition of 
analyticity at $\lambda=0$, so $\exists$ $H_0$ limit, we can calculate anything in the deformed theory from the undeformed theory, 
in particular the correlations functions are found by a general formula. Moreover, obviously (with the extra analyticity assumption, 
needed for the case of an infinite dimensional Hilbert space) we have the same spectrum of eigenstates for the two Hamiltonians. 

But note that this does not mean that the Lagrangians are also related, $L_\lambda\neq g(\lambda, L_0)$!! We can easily see 
this in the formula for $L$ (\ref{Lagnew}), in the nontrivial case of $V\neq 0$. Reversely, if $L_\lambda=f(\lambda, L_0)$, 
it doesn't mean that $H_\lambda=f(\lambda,H_0)$. 

For the DBI example = $T\bar T$ deformation of the $V=0$ case in 2 dimensions, 
\be
{\cal L}=\frac{1}{\lambda}\left[1-\sqrt{1+\lambda (-\dot\phi^2+\phi'^2)}\right]\;,
\ee
so 
\be
p=\frac{\dot\phi}{\sqrt{1+\lambda(-\dot\phi^2+\phi'^2)}}\;,
\ee
so 
\be
H=\frac{1}{\lambda}\left[-1+\sqrt{(1+\lambda p^2)(1+\lambda \phi'^2)}\right]\;,
\ee
so it is not a function of $H_0=\frac{1}{2}(p^2+\phi'^2)$.



\section{Conclusions and discussion}

In this paper we have considered $T\bar T$ deformations in the context of holography, and more 
specifically in the context of the pp wave correspondence. 

We have applied the TsT procedure of 
obtaining the gravity dual to a single trace $T\bar T$ deformation in the $AdS_3\times S^3\times T^4$
case to the $AdS_5\times S^5$ case, and used the Penrose limit to understand it, proposing that the 
deformation corresponds to some dipole theory, probably a noncommutative one. 

Reversely, we have considered the $T\bar T$ deformation of the Penrose limit of $AdS_5\times S^5$, in two 
ways. We have discretized the $T\bar T$ deformation of the worlsheet string on the pp wave, though 
the corresponding spin chain looks complicated. We have instead considered the $T\bar T$ 
deformation of the 
quantum mechanical model obtained from the discretization of the string Hamiltonian on the pp wave, 
known to correspond to the spin chain one in ${\cal N}=4$ SYM. Based on the corresponding symmetry
algebra of the deformed pp wave theory, we have argued that the deformation can be understood 
within ${\cal N}=4$ SYM as  deformation of the (BMN) sector within it, but not of the theory, obtaining 
moreover an equivalent sector. 

There are many open questions left for further work. The exact nature of the deformation to ${\cal N}=4$
SYM dual to the TsT transformed $AdS_5\times S^5$ is not yet clear. The spin chain obtained 
by discretizing the $T\bar T$ deformed worlsheet string for the pp wave has no good interpretation yet. 
One can try to use the same methods considered here in other holographic cases for dimensions $d>2$,
for instance the $d=3$ case of the ABJM model, as well as more realistic cases 
like Klebanov-Witten, Klebanov-Strassler, Maldacena-Nunez, Maldacena-Nastase. In particular holographic pp wave theories associated with confining gauge dynamics analyzed in \cite{Gimon:2002nr}.
Finally, we have only considered the bosonic part of the string worldsheet theory on the pp wave; it 
would be interesting to see what happens when we add the fermions, in all the cases considered. 
This is in particular interesting, since one can construct supersymmetric $T\bar T$ deformations, 
and the deformation appears as a supersymmetric descendant \cite{Baggio:2018rpv,Chang:2018dge}.

\section*{Acknowledgements}
We would like to thank Thiago Araujo for collaboration at the early stages of this project, and to
Stefano Negro for discussions and comments on the manuscript.

The work of HN is supported in part by  CNPq grant 301491/2019-4 and FAPESP grants 2019/21281-4 
and 2019/13231-7. HN would also like to thank the ICTP-SAIFR for their support through FAPESP grant 2016/01343-7. The work of J.S was supported  by a center of excellence of the Israel Science Foundation (grant number 2289/18).
 
\appendix

\section{Review of derivation of $T\bar T $ deformation of general quantum mechanical model} 

Here we review the derivation in \cite{Gross:2019ach} of the $T\bar T$ deformation of general quantum mechanics systems, 
via dimensional reduction, assuming the $AdS_3/CFT_2$. 

In the 2 dimensional CFT, it is assumed that the flow equation is 
\be
\frac{\d S_E(\lambda)}{\d \lambda}=\int d^2x \sqrt{\gamma} 8T\bar T\;,
\ee
and that along the flow, we have the equation (following from holography of the $T\bar T$ deformation following
\cite{McGough:2016lol})
\be
T_\mu^\mu=-16\lambda T\bar T =-2T(T_{ij} T^{ij}-(T_i^i)^2).
\ee

The $AdS_3/CFT_2$ gravity dual of BTZ-BH type is 
\be
ds^2=(r^2-r_+^2)d\tau^2+\frac{dr^2}{r^2-r_+^2}+r^2d\phi^2\rightarrow_{bd.} r^2(d\tau^2+d\phi^2).
\ee

Then the condition on the flow is 
\be
T_\mu^\mu=T_\phi^\phi+T_\tau^\tau=-2\lambda(T_{ij}T^{ij}-(T_i^i)^2)\;,
\ee
and 
\bea
T_{ij} T^{ij}-(T_i^i)^2&=&T_{\tau\tau}T^{\tau\tau}+T_{\phi\phi}T^{\phi\phi}+2T_{\tau\phi}T^{\tau\phi}-(T_\tau^\tau+T_\phi^\phi)^2\cr
&=& 2T_{\tau\phi}T^{\tau\phi}+2T_\tau^\tau T_\phi^\phi\;,
\eea
so the condition along the flow becomes 
\be
T_\phi^\phi=\frac{T^{\tau_\tau}+4\lambda T_{\tau\phi}T^{\tau\phi}}{4\lambda T_\tau^\tau-1}\;,
\ee
and, given that $T_\mu^\mu=T_\phi^\phi+T_\tau^\tau=-16\lambda T\bar T$ on the flow, we have 
\be
8T\bar T=\frac{T_\phi^\phi+T_\tau^\tau}{-2\lambda}=\frac{(T_\tau^\tau)^2+T_{\tau\phi}T^{\tau\phi}}{1/2-2\lambda T_\tau^\tau}\Rightarrow
\frac{\d S_E(\lambda)}{\d \lambda}=\int d^2x \sqrt{\gamma}\frac{(T_\tau^\tau)^2+T_{\tau\phi}T^{\tau\phi}}{1/2-2\lambda T_\tau^\tau}.
\ee

But, since $\langle T_{\tau\phi}\rangle=\langle T^{\tau\phi}\rangle =iJ$ and $\langle T_\tau^\tau\rangle=E$, 
by which we mean more precisely integral over a circle of radius 1, $\int T_{\tau\phi}=\int T^{\tau\phi}=iJ$ and $\int T_\tau^\tau=E$
(and implicit factorization of the square) besides the VEV, putting $\gamma_{\a\b}=\delta_{\a\b}$, peeling off the time integral in $S_E$, 
we obtain the equation for the energy eigenstates,
\be
\frac{\d E(\lambda)}{\d \lambda}=\frac{E^2-J^2}{1/2-2\lambda E}\Rightarrow E(\lambda)=\frac{1}{4\lambda}\left(1-\sqrt{1-8\lambda
E_0+16\lambda^2J^2}\right)\;,
\ee
the usual solution.

Dimensional reduction proceeds as follows: put $T_{\phi\tau}=T^{\tau\phi}=0$ and $T_\tau^\tau\equiv T$, to obtain 
\be
\frac{\d S_E}{\d\lambda}=\int d\tau \frac{T^2}{1/2-2\lambda T}.
\ee

Then, again peeling of $d\tau$ and using $\langle T\rangle=E$ (VEV) and factorization of the square, we obtain the 
equation for the energy eigenstates, 
\be
\frac{\d E(\lambda)}{\d \lambda}=\frac{E^2}{1/2-2\lambda E}\Rightarrow E(\lambda)=\frac{1-\sqrt{1-8\lambda E_0}}{4\lambda}\;,
\ee
or, equivalently as we see, just put $J=0$ in the 2-dimensional result. 

 Note that if we take $\lambda\rightarrow -\lambda$, we obtain 
\be
\frac{\d E(\lambda)}{\d \lambda}=-\frac{E^2}{1/2+2\lambda E}\Rightarrow E(\lambda)=-\frac{1-\sqrt{1+8\lambda E_0}}{4\lambda}.
\ee

See the discussion in the main text with respect to the signs in $H$ and $L$.

Now, we assume then that the same relation is true for the Hamiltonians (since it is valid for all its eigenstates, and the Hamiltonian 
is a hermitian operator), so 
\be
H(\lambda)=\frac{1}{4\lambda}\left(1-\sqrt{1-8\lambda H_0}\right)
=\frac{1}{4\lambda}\left(1-\sqrt{1-8\lambda\left(\sum_i\frac{p_i^2}{2}+V(\{q_i\})\right)}\right).\label{defHam}
\ee

\bibliography{TTbar}
\bibliographystyle{utphys}

\end{document}